\documentclass[12pt]{article}
\usepackage{putex}
\pdfoutput=1
\usepackage{graphicx}
\usepackage{epstopdf}
\usepackage{enumerate}
\usepackage{cite}

\newcommand{\norm}[1]{\lVert #1 \rVert}
\newcommand{\abs}[1]{\lvert #1 \rvert}

\newcommand{\ord}{\mathcal {O}}

\newcommand {\be} {\begin {equation}}
\newcommand {\ee} {\end {equation}}

\newcommand {\bes} {\begin {equation*}}
\newcommand {\ees} {\end {equation*}}

\newcommand{\es}[2] {\begin{equation} \label{#1} \begin{split} #2 \end{split} \end{equation}}

\newcommand{\CP}{\mathbb{CP}}
\newcommand{\Z}{\mathbb{Z}}
\newcommand{\R}{\mathbb{R}}
\newcommand{\C}{\mathbb{C}}

\begin{document}

\preprint{PUPT-2336}

\institution{PU}{Joseph Henry Laboratories$^1$ and Center for Theoretical Science,$^2$\cr
~~~~~~~~~~~~~~Princeton University, Princeton, NJ 08544}

\title{Membranes with Topological Charge and AdS${}_4$/CFT${}_3$ Correspondence}

\authors{Igor R. Klebanov,$^{1,2}$ Silviu S. Pufu,$^1$ and Tiberiu Tesileanu$^1$}

\abstract{If the second Betti number $b_2$ of a Sasaki-Einstein manifold $Y^7$ does not vanish, then M-theory on $AdS_4\times Y^7$ possesses ``topological'' $U(1)^{b_2}$ gauge symmetry. The corresponding Abelian gauge fields come from three-form fluctuations with one index in $AdS_4$ and the other two in $Y^7$.  We find black membrane solutions carrying one of these $U(1)$ charges. In the zero temperature limit, our solutions interpolate between $AdS_4\times Y^7$ in the UV and $AdS_2\times \R^2\times \text{squashed } Y^7$ in the IR\@.  In fact, the $AdS_2 \times \R^2 \times \text{squashed }Y^7$ background is by itself a solution of the supergravity equations of motion.  These solutions do not appear to preserve any supersymmetry.  We search for their possible instabilities and do not find any. We also discuss the meaning of our charged membrane backgrounds in a dual quiver Chern-Simons gauge theory with a global $U(1)$ charge density. Finally, we present a simple analytic solution which has the same IR but different UV behavior. We reduce this solution to type IIA string theory, and perform T-duality to type IIB\@. The type IIB metric turns out to be a product of the squashed $Y^7$ and the extremal BTZ black hole.  We discuss an interpretation of this type IIB background in terms of the $(1+1)$-dimensional CFT on D3-branes partially wrapped over the squashed $Y^7$.
}

\date{April 2010}

\maketitle

\tableofcontents

\section{Introduction}

Over the past few years, considerable effort has been devoted to using the anti-de Sitter / conformal field theory (AdS/CFT) correspondence \cite{Maldacena:1997re,Gubser:1998bc,Witten:1998qj} for studying strongly coupled field theories at non-vanishing chemical potential for some conserved global charge (see \cite{Hartnoll:2009sz, Herzog:2009xv} for reviews). A global symmetry of a $(p+1)$-dimensional conformal field theory is mapped to a gauge symmetry in $(p+2)$-dimensional anti-de Sitter space. Therefore, properties of a conformal field theory at finite chemical potential $\mu$ and temperature $T$ are encoded in charged $p$-brane solutions that are asymptotic to $AdS_{p+2}\times Y$, where $Y$ is an Einstein space.

An interesting class of AdS/CFT dualities involves Sasaki-Einstein spaces $Y$ which lead to backgrounds preserving eight supercharges. Type IIB backgrounds of the form $AdS_5\times Y^5$ are therefore dual to ${\cal N}=1$ superconformal gauge theories in four space-time dimensions, while M-theory backgrounds $AdS_4\times Y^7$ are dual to ${\cal N}=2$ superconformal gauge theories in three space-time dimensions \cite{Kehagias:1998gn,Klebanov:1998hh,Acharya:1998db}. These theories possess $U(1)_R$ symmetry that in supergravity is realized as an isometry of $Y$.  In an effective $(p+2)$-dimensional description, the charged black branes are described by the well-known Reissner-Nordstr\" om AdS (RNAdS) backgrounds. One typically finds, however, that as the temperature divided by the chemical potential is reduced, such a charged $p$-brane solution is not the thermodynamically-preferred phase of the theory \cite{Cvetic:1999ne,Chamblin:1999tk,Cvetic:1999rb}; it becomes unstable towards developing charged ``hair'' \cite{Gubser:2008px}. Typically, when an R-charged $p$-brane is embedded into string or M-theory, there are R-charged fields that condense close to the black hole horizon, thus breaking the $U(1)$ gauge symmetry spontaneously \cite{Denef:2009tp,Gubser:2009qm,Gauntlett:2009dn,Gauntlett:2009bh}.  The corresponding symmetry breaking in the field theory has been used to model superconductivity \cite{Hartnoll:2008vx} or superfluidity \cite{Herzog:2008he} in a strongly coupled CFT\@.

For other classes of applications, however, it is desirable that the symmetric phase is stable down to very low (or even vanishing) temperature \cite{Lee:2008xf,Liu:2009dm, Cubrovic:2009ye,Faulkner:2009wj,Rey:2008zz}. If so, then there exists a quantum critical phase described by the $AdS_2\times \R^p$ extremal near-horizon region of the RNAdS background. In view of their potential low temperature instabilities, embedding such backgrounds into string theory or M-theory poses an interesting challenge. Luckily, in addition to the isometries of the Sasaki-Einstein space $Y$, the $AdS_{p+2}\times Y$ backgrounds may have some non-R $U(1)$ symmetries. The corresponding gauge fields in $AdS_{p+2}$ arise due to the non-trivial topology of $Y$.  The number of such topological $U(1)$ symmetries is given by the second Betti number, $b_2$, of the internal space $Y$. In general, the $n$th Betti number $b_n$ equals the number of linearly independent harmonic $n$-forms on the manifold $Y$, each of these forms representing a generator of the de Rham cohomology $H^n(Y)$.
In the $10$-d examples from type IIB supergravity, the space $Y$ is five-dimensional and by the Poincar\'e duality $b_2 = b_3$;  in the $11$-d examples, the space $Y$ is seven-dimensional, and the Poincar\'e duality implies $b_2 = b_5$.

The connection between topology and supergravity fluctuations comes about as follows \cite{DAuria:1984vv,Klebanov:1999tb,Ceresole:1999zs}. In $AdS_{p+2} \times Y$ compactifications, harmonic forms on $Y$ are all that is needed to construct a consistent linearized set of fluctuations that includes massless gauge fields in $AdS_{p+2}$, one gauge field for each of the linearly independent harmonic forms.  Let us examine in more detail the $AdS_5 \times Y^5$ compactifications of type IIB SUGRA\@.  Denoting by $\omega_3^{(i)}$ the $b_3$ linearly independent harmonic three-forms on $Y^5$ and by $\omega_2^{(i)}$ their Hodge duals (which are harmonic two-forms), one can consider fluctuations in the type IIB SUGRA R-R sector of the form
\es{Top3Currents}{
\delta C_4 = \sum_{i = 1}^{b_2} \left[ A^{(i)} \wedge \omega_3^{(i)} + \tilde B^{(i)} \wedge \omega_2^{(i)} \right] \,,
}
where $A^{(i)}$ are one-forms in $AdS_5$ and $\tilde B^{(i)}$ are two-forms, and one can work only to linear order in the $A^{(i)}$ and $\tilde B^{(i)}$. The fluctuations \eqref{Top3Currents} are intrinsically non-geometric because they don't involve the metric. The two-forms $\tilde B^{(i)}$ are related to the one-forms $A^{(i)}$ through the duality relation $d\tilde B^{(i)} = *_5 dA^{(i)}$ in $AdS_5$, so one can consider only the $A^{(i)}$ to be fundamental variables.  This duality relation implies $d *_5 dA^{(i)} = 0$, which shows that the $A^{(i)}$ are massless gauge fields from an effective $AdS_5$ perspective.   A careful examination of the type IIB supergravity action shows that to linear order in $A^{(i)}$ no other fluctuations mix with $\delta C_4$ in \eqref{Top3Currents}. The Bianchi identity $dF_5 = 0$ is satisfied automatically, and the self-duality constraint for $F_5 = dC_4$, $F_5 = * F_5$,  is satisfied when $\omega_3^{(i)}$ and its Hodge dual $\omega_2^{(i)}$ are closed, or in other words when $\omega_3^{(i)}$ and $\omega_2^{(i)}$ are harmonic. To summarize, the harmonicity of $\omega_3^{(i)}$ and $\omega_2^{(i)}$ is a crucial ingredient in making \eqref{Top3Currents} a consistent linearized ansatz and for the existence of the gauge fields $A^{(i)}$ in $AdS_5$.  In turn, the existence of the harmonic forms $\omega_3^{(i)}$ and $\omega_2^{(i)}$ is determined only by the topology of $Y^5$.

A well-known example of this kind is $Y^5=T^{1,1}$ with $b_3=1$, so there is only one topological $U(1)$ symmetry.   This symmetry is referred to as baryonic because in the dual $SU(N)\times SU(N)$ superconformal gauge theory \cite{Klebanov:1998hh} the only operators that are charged under it are the ``di-baryons,'' which involve products of at least $N$ bi-fundamental fields \cite{Gubser:1998fp}.  The three-brane backgrounds asymptotic to $AdS_5\times T^{1,1}$ that are charged under this topological baryonic symmetry were constructed in \cite{Herzog:2009gd}. These solutions contain a novel nearly conformal infrared region. In contrast to the RNAdS solution, this infrared region is not simply of the direct product form $AdS_2\times \R^3\times \text{squashed }T^{1,1}$;  it differs from it by slowly varying warp factors that are powers of the logarithm of the $AdS_2$ radius.  In the dual IR field theory this should be interpreted as a logarithmic RG flow.\footnote{At low temperatures, this background is not perfectly stable.  There is a metastability associated with the ``Fermi seasickness'' of \cite{Hartnoll:2009ns}, i.e.~the tunneling of space-time filling D3-branes to $AdS_5$ infinity (see also \cite{Yamada:2007gb, Yamada:2008em}).  In the dual field theory this phenomenon may be due to the presence of charged massless scalar fields.}

In this paper we extend the study of topological charge densities to a class of M-theory Freund-Rubin compactifications, $AdS_4 \times Y^7$, where $Y^7$ is a seven-dimensional Sasaki-Einstein manifold with $b_2>0$.  The general idea behind obtaining $b_2$ gauge fields $A^{(i)}$ in $AdS_4$ using the $b_2$ linearly-independent harmonic forms on $Y^7$ \cite{DAuria:1984vv} is the same as for $AdS_5 \times Y^5$, but the details are somewhat different.  Denoting by $\omega_2^{(i)}$ the $b_2$($ = b_5$) linearly independent harmonic two-forms on $Y^7$ and by $\omega_5^{(i)}$ their seven-dimensional Hodge duals (which in this case are harmonic five-forms on $Y^7$), one can consider the following consistent set of linearized fluctuations of eleven-dimensional supergravity:
\es{Top2Currents}{
\delta A_3 = \sum_{i = 1}^{b_2} A^{(i)} \wedge \omega_2^{(i)}\ , \qquad
\delta A_6 = \sum_{i = 1}^{b_2} \tilde A^{(i)} \wedge \omega_5^{(i)}\ , \qquad
d\tilde A^{(i)} = *_4 d A^{(i)} \,,
}
where $A^{(i)}$ and $\tilde A^{(i)}$ are one-forms in $AdS_4$.  Similarly to the $AdS_5 \times Y^5$ case, the duality relation  $dA_6 = *dA_3$ requires that the fields $A^{(i)}$ and $\tilde A^{(i)}$ should be related to each other through $d\tilde A^{(i)} = *_4 d A^{(i)}$, and that $\omega_2^{(i)}$ and $\omega_5^{(i)}$ should be harmonic forms.  The relation $d\tilde A^{(i)} = *_4 d A^{(i)}$ implies that both $A^{(i)}$ and $\tilde A^{(i)}$ satisfy the equation of motion for a gauge field, $d*_4 dA^{(i)} = d*_4 d \tilde A^{(i)} = 0$. For each $i$, there are two different boundary conditions in $AdS_4$ which correspond to treating either $A^{(i)}$ or $\tilde A^{(i)}$ as the fundamental variable \cite{Witten:2003ya,Imamura:2008ji}.  The two possible conserved charges, electric and magnetic, map in the dual gauge theory to global charge density and magnetic field, respectively \cite{Hartnoll:2007ip, Herzog:2007ij}.  For our purposes, this choice corresponds to allowing either the wrapped M2-branes or the wrapped M5-branes.   We will comment on the dual field theory interpretation of the $AdS_4 \times Y^7$ backgrounds, and the meaning of this choice, in section~\ref{FIELDTHEORY}. The above discussion shows that in the M-theory case the supergravity fluctuation spectrum around $AdS_4 \times Y^7$ contains $b_2$ independent gauge fields whose existence relies on the existence of harmonic two- and five-forms on $Y^7$.

In this paper we will consider Sasaki-Einstein spaces $Y^7$ which are principal $U(1)$ bundles over a direct product of two K\"ahler-Einstein spaces, $V_1$ and $V_2$. In this case, there exists a universal harmonic two-form $\omega_2$ (or, equivalently, a universal harmonic five-form $\omega_5$) that we exhibit in the next section.  We will construct two-brane solutions electrically charged under the corresponding gauge field $A$ coming from $\delta A_3$.\footnote{An ansatz for magnetically charged solutions was set up in \cite{Herzog:2009gd}, but seems to lead to backgrounds singular in the IR\@.}  As in the solutions of \cite{Herzog:2009gd}, several warp factor functions enter our consistent non-linear ansatz. We derive a system of coupled ODEs for these functions and solve them numerically to find the backgrounds for various values of $T/\mu$. The warp factors turn out to stabilize to finite nonzero values at the horizon in the zero-temperature limit, producing an $AdS_2\times \R^2\times \text{squashed }Y^7$ throat region that is also a solution to $11$-d supergravity. We find numerically the extremal background interpolating between this throat region in the IR and $AdS_4\times Y^7$ in the UV\@.  We also find an analytic solution with the same IR but different UV behavior. A possible instability associated with condensation of charged fields would manifest itself in wrapped probe M2-branes being repelled from the horizon. However, using the M2-brane world volume action, we show quite generally that such an instability does not occur.  We make some simple checks of stability against condensation of neutral scalar fields, and we again find no instabilities. We also study the potential for a probe space-time filling M2-brane and prove that it vanishes at $T=0$. Hence, there is no brane nucleation instability, and our solution seems to be a good candidate for embedding the $AdS_2\times \R^2$ IR behavior into M-theory.

The rest of this paper is organized as follows.  In section~\ref{TRUNCATION} we describe the eleven-dimensional ansatz and construct the charged black membranes numerically at nonzero temperature and chemical potential.  In section~\ref{EXTREMAL} we find the zero-temperature limit of our backgrounds and show that
the $AdS_2 \times \R^2 \times \text{squashed }Y^7$ throat by itself satisfies the $11$-d supergravity equations of motion.  We
also present a similar analytic solution with different large $r$ behavior. In section~\ref{M2FORCE} we compute the potential for the charged objects in the theory---the M2-branes.  In section~\ref{FIELDTHEORY} we discuss an interpretation of our results in the dual quiver Chern-Simons gauge theories. The wrapped M2-branes are dual to operators containing non-diagonal magnetic fluxes, and we comment on their fractional statistics. In section \ref{DUALITIES} we use string dualities to map our analytic solution to one in type IIB theory, and find that the type IIB metric is a product of the squashed $Y^7$ and the extremal BTZ black hole (the one that has the minimum mass for a given angular momentum in $AdS_3$) \cite{Banados:1992wn, Banados:1998gg}. The Appendices contain some further stability checks and constructions of the two-cycles in $Y^7$.

\section{A universal consistent truncation}
\label {TRUNCATION}

Let us consider a seven-dimensional Einstein space $Y^7$ that can be written as a $U(1)$ fiber bundle over a direct product of two K\"ahler-Einstein spaces, $V_1$ and $V_2$.  The spaces $Y^7$, $V_1$, and $V_2$ could be manifolds or, more generally, orbifolds.  The product $V_1 \times V_2$ must describe a space of real dimension six, or complex dimension three, so without loss of generality we assume that $V_1$ and $V_2$ have complex dimensions two and one, respectively.  In section~\ref{GEOMETRY} we first show explicitly that all the spaces $Y^7$ with the property mentioned above admit a universal harmonic two-form which can be used to construct a massless gauge field in $AdS_4$ \eqref{Top2Currents}, and then give a non-linear consistent truncation of eleven-dimensional supergravity that allows us to construct black membrane solutions with topological charge.  In section~\ref{EXAMPLES} we give examples of spaces $Y^7$.  Section~\ref{THERMO} is concerned with examining the thermodynamic properties of the charged black branes at nonzero temperature and charge density.  Lastly, in section~\ref{NUMERICS} we construct these black branes numerically.

\subsection {The eleven-dimensional background}
\label {GEOMETRY}

Quite generally, the Einstein metric on the space $Y^7$ can be written as
\es{YMetric}{
ds_Y^2 = ds_{V_1}^2 + ds_{V_2}^2 + \left(d \psi + \sigma_1 + \sigma_2 \right)^2 \,,
}
where each of the connection one-forms $\sigma_i$ is a pull-back of a locally-defined one-form on $V_i$.  It is convenient to normalize this metric so that in a vielbein basis $R_{ab} = 6 \delta_{ab}$.  The Einstein condition for $Y^7$ implies both that
\es{sigmaKahler}{
d \sigma_i = 2 \omega_i \,,
}
where $\omega_i$ is the K\"ahler form on $V_i$, and that the Einstein metric on $V_i$ should be normalized so that the curvature two-form $R_i$ satisfies $R_i = 8 \omega_i$.  In this normalization, the range of $\psi$ depends on the first Chern class of the fibration;  see Appendix~\ref{METRICS} for more details.

The spaces $Y^7$ admit a universal harmonic two-form given by
\es{HarmonicTwoForm}{
\omega \equiv \omega_1 - 2 \omega_2 \,.
}
To see that this form is harmonic, it is helpful to pass to a vielbein basis where, in a small enough coordinate patch, $\omega_1 = e_1 \wedge e_2 + e_3 \wedge e_4$, $\omega_2 = e_5 \wedge e_6$, and $d \psi + \sigma_1 + \sigma_2 = e_7$.  In this basis, the volume form on $Y^7$ is just $\vol_Y = e_1 \wedge e_2 \wedge e_3 \wedge e_4 \wedge e_5 \wedge e_6 \wedge e_7$.   The Hodge dual of $\omega$ can then be computed to be
\es{HodgeDuals}{
*_Y\,  \omega = \omega_1 \wedge (\omega_2 - \omega_1) \wedge (d \psi + \sigma_1 + \sigma_2) \,.
}
Using \eqref{sigmaKahler}, \eqref{HodgeDuals}, and the fact that both $\omega_1$ and $\omega_2$ are closed, one can show that $d\omega = d*_Y \omega = 0$, so $\omega$ is indeed harmonic. Note also that $J\wedge *_Y\,  \omega=0$, where
\es{KahlerForm}{
J \equiv \omega_1 + \omega_2
}
is the K\"ahler form on $V_1 \times V_2$.

One can use the space $Y^7$ and $\omega$ to construct a charged black hole solution to the eleven-dimensional supergravity equations of motion as follows.\footnote{Throughout this paper we follow the conventions of \cite{Polchinski:1998rr} for the supergravity actions.}  The eleven-dimensional metric is a warped product of a non-compact four-dimensional space $M$ and a squashed version of \eqref{YMetric}:
\es{11dMetric}{
ds^2 =  e^{-7\chi/2} ds_M^2 + 4 L^2 e^\chi \left[
e^{\eta_1} ds_{V_1}^2 + e^{\eta_2} ds_{V_2}^2
+ e^{-4 \eta_1 - 2 \eta_2} (d \psi + \sigma_1 + \sigma_2)^2  \right] \,,
}
where the scalar fields $\chi$, $\eta_1$, and $\eta_2$ are functions only of the coordinates on $M$.  In fact, we will only look for static solutions that are rotationally symmetric in two of the four non-compact directions, and we write the metric on $M$ in the form
\es{MetricM}{
ds_M^2 = -g e^{-w} dt^2 + {r^2 \over L^2} \left[(dx^1)^2 + (dx^2)^2 \right]
+ {dr^2 \over g} \,,
}
where $g$, $w$, $\chi$, $\eta_1$, and $\eta_2$ depend only on $r$.

In addition to the metric, we need to specify the four-form $F_4$:
\es{F4M2}{
F_4 &= -{3\over L} e^{-{21\over 2} \chi} \vol_M
-8  Q L^3 {e^{-{w \over 2}  - {3 \over 2} \chi} \over r^2} dt \wedge dr \wedge
\left( e^{ 2 \eta_1} \omega_1 - 2 e^{2 \eta_2} \omega_2 \right)    \,,
}
where $Q$ is a constant related to the charge of the black hole, and the orientation of $M$ is given by
\es{VolMDef}{
 \vol_M \equiv {r^2 \over L^2} e^{-{1 \over 2} w} dt \wedge dx^1 \wedge dx^2 \wedge dr \,.
}
Its Hodge dual, $F_7$, has the form
\es{GotF7}{
F_7 = 384\, L^6 \vol_Y
+64\, Q L^4 dx^1 \wedge dx^2 \wedge (*_Y \omega) \,,
}
with $*_Y \omega$ defined as in \eqref{HodgeDuals}.  When $Q$ is small, the $11$-d equations of motion imply that $\eta_1$, $\eta_2$, and $\chi$ are of order ${\cal O}(Q^2)$, so to linear order in $Q$, equations \eqref{F4M2}--\eqref{GotF7} take the form \eqref{Top2Currents} with the gauge fields $A$ and $\tilde A$ having only electric and only magnetic components, respectively.

The effective one-dimensional Lagrangian describing the consistent truncation \eqref{MetricM}--\eqref{GotF7} is
\es{EffectiveLag}{
\mathcal L &= {r^2 \over L^2} e^{-{w \over 2}} \biggl[
{63 g \over 8} \chi'^2 + {g \over 2} (2\eta_1'^2 + \eta_2'^2)
+ g (2 \eta_1' + \eta_2')^2
+ {2 g \over r} w' - {2 \over r} g' - {2 g \over r^2} + V_Q + V_s \biggr] \,,
}
where
\es{GotPotentials}{
V_Q &= {4 L^2 \over r^4} e^{- {3 \over 2} \chi} \left(e^{2 \eta_1}
+ 2 e^{2 \eta_2} \right) Q^2 \,, \\
V_s &= {9 \over 2 L^2} e^{-{21 \over 2} \chi}
- {4 \over L^2} e^{-{9 \over 2} \chi} \left( 2 e^{-\eta_1} + e^{-\eta_2} \right)
+ {1 \over 2 L^2} e^{-2 (2 \eta_1 + \eta_2 )-{9 \over 2} \chi}
\left[2 e^{-2 \eta_1} + e^{-2 \eta_2} \right] \,.
}
This Lagrangian needs to be supplemented by the zero-energy constraint
\es{ZeroEnergy}{
{2 \over r} g'  - g \left[ {63 \over 8} \chi'^2 + {1\over 2} (2\eta_1'^2 + \eta_2'^2)
+ (2 \eta_1' + \eta_2')^2
+ {2 \over r} w'  - {2 \over r^2} \right] + V_Q + V_s = 0 \,.
}
The scalar potential $V_s$ agrees with the one derived in \cite{Ahn:1999zy} for the particular case where the Sasaki-Einstein manifold $Y^7$ is $Q^{1, 1, 1}$.

\subsection{Examples}
\label{EXAMPLES}

Examples of spaces $Y^7$ satisfying the requirements of the previous section are some regular Sasaki-Einstein manifolds and orbifolds thereof. A Sasaki-Einstein manifold $Y$ is a compact Riemannian manifold whose metric cone is Calabi-Yau.  A Sasaki-Einstein manifold can be described as a principal $U(1)$ bundle over a K\"ahler-Einstein base $V$, which in general cannot be written as a product $V_1 \times V_2$ as in the previous section.  A Sasaki-Einstein manifold is called regular if the fibers all close and have the same length.

There aren't many examples of regular Sasaki-Einstein manifolds in seven dimensions.  An exhaustive list is given by \cite{Friedrich:1990zg}:
\begin{enumerate}[I.]
\item Regular $SE_7$ where the base $V$ cannot be written as a product $V_1 \times V_2$:
\begin {itemize}
\item $S^7$, which is a $U(1)$ fibration over $\CP^3$.
\item $N^{0, 1, 0}$, which is a $U(1)$ fibration over the flag manifold $F(1, 2)$.
\item $V_{5, 2}$, which is a $U(1)$ fibration over the Grassmanian manifold $G_{5, 2}$.
\end {itemize}
\item Regular $SE_7$ where $V = V_1 \times V_2$:
\label {Interesting}
\begin {itemize}
\item $Q^{1, 1, 1}$, which is a $U(1)$ fibration over $\CP^1 \times \CP^1 \times \CP^1$.
\item $Q^{2, 2, 2}$, which is a $\Z_2$ orbifold of $Q^{1, 1, 1}$ and a $U(1)$ fibration over $\CP^1 \times \CP^1 \times \CP^1$ also.  It differs from $Q^{1, 1, 1}$ in that the length of the fiber is shorter by a factor of two.
\item $M^{1, 1, 1}$, which is a $U(1)$ fibration over $\CP^2 \times \CP^1$.
\item Spaces which we will call ${\cal P}_n$ that are appropriate $U(1)$ fibrations over $dP_n \times \CP^1$, $3 \leq n \leq 8$, where $dP_n$ is the $n$th del Pezzo surface constructed by blowing up $\CP^2$ at $n$ generic points.
\end {itemize}
\end{enumerate}
From now on we will only be interested in the second group of examples listed above for which the base of the $U(1)$ fibration can be written as a direct product of two K\"ahler-Einstein manifolds.  Indeed, $V = V_1 \times V_2$ was a necessary ingredient for constructing the consistent truncation of eleven-dimensional supergravity presented in section~\ref{GEOMETRY}.

In addition to the regular Sasaki-Einstein spaces we just described, one can also consider their orbifolds.  While the regular Sasaki-Einstein spaces under \eqref{Interesting} all preserve eight supercharges, their orbifolds generically break all SUSY.

\subsection{Thermodynamics}
\label {THERMO}

\subsubsection {Boundary conditions}
Before we calculate thermodynamic quantities, we need to discuss the boundary conditions one should impose on the solutions to the equations of motion following from \eqref{EffectiveLag}--\eqref{ZeroEnergy}.  At large $r$, these solutions should asymptote to $AdS_4 \times Y^7$, so
\be
\label {BoundaryConds}
\begin {aligned}
w \to 0\,,  \qquad \chi &\to 0\,, \qquad \eta_1 \to 0\,, \qquad \eta_2\to 0\,,\\
g &= \frac {r^2} {L^2} + \ord (L/r)\,.
\end {aligned}
\ee
Generically, there will be an event horizon at some $r = r_h$ where $g$ vanishes.  The remaining boundary conditions come from requiring regularity of all the fields at $r = r_h$.

Let us examine the boundary conditions \eqref{BoundaryConds} more carefully.  From the asymptotic form of the equations at large $r$ we find that there is only one possible behavior for $\chi$ consistent with \eqref{BoundaryConds}:  $\chi \sim 1/r^6$.  The gauge theory operator dual to $\chi$ has conformal dimension $\Delta_\chi = 6$, because in general the bulk field dual to a scalar operator of dimension $\Delta$ behaves at large $r$ as $r^{-\Delta}$ if no sources for that operator are turned on.  A similar asymptotic analysis shows that the fields $\eta_{1,2}$ break up into the combinations
\newcommand {\meta} {\tilde \eta}
\newcommand {\mlambda} {\tilde \lambda}
\be
\meta = \frac {2\eta_1 + \eta_2} 3 \,, \qquad \mlambda =
\frac {\eta_1 - \eta_2} 3
\ee
that have definite scaling dimensions at large $r$.  While for $\meta$ there is only one possible large $r$ behavior consistent with \eqref{BoundaryConds}, $\meta \sim 1/r^4$, corresponding to $\Delta_{\meta} = 4$, $\mlambda$ generically behaves as a linear combination of $1/r$ and $1/r^2$ with arbitrary coefficients, both of these behaviors being consistent with $AdS_4 \times Y^7$ asymptotics.  One then has a choice of boundary conditions where either $\Delta_{\mlambda} = 1$ and the coefficient of $1/r^2$ is required to vanish, or $\Delta_{\mlambda} = 2$ and the coefficient of $1/r$ is required to vanish \cite{Klebanov:1999tb}. In this paper we choose the latter boundary condition on $\mlambda$. With this choice, the equations obtained from the Lagrangian~\eqref {EffectiveLag} subject to the zero-energy constraint~\eqref{ZeroEnergy} and the other boundary conditions described above can be solved by a power series expansion at large $r$. The first few terms in the expansion are given below:
\be
\label {BoundaryExp}
\begin {aligned}
w &= \ord (L^4/r^4) \,,\\
g &= \frac {r^2} {L^2} + \frac {g_1 L} r + \ord (L^2/r^2)\,,\\
\chi &= \frac {2 L^4} {49 r^4} \bigl (2Q^2 - 3 \lambda_2^2\bigr) +
\frac {\chi_6 L^6} {r^6} + \ord (L^7/r^7) \,,\\
\eta_1 &= \frac {\lambda_2 L^2} {r^2} - \frac {4 L^4} {35 r^4}
\bigl (2Q^2 - 3\lambda_2^2 \bigr) \log \frac rL -
\frac {L^4} {r^4}\biggl(\frac 23Q^2 + \eta_4\biggr) + \ord (L^6/r^6) \,,\\
\eta_2 &= -\frac {2\lambda_2 L^2} {r^2} - \frac  {4 L^4} {35 r^4}
\bigl (2Q^2 - 3 \lambda_2^2\bigr) \log \frac rL +
\frac {L^4} {r^4} \biggl(\frac 43 Q^2 + \eta_4\biggr) + \ord (L^5/r^5) \,.
\end {aligned}
\ee
All higher order terms are determined in terms of $g_1$, $\chi_6$, $\eta_4$, $\lambda_2$, and $Q$.

\subsubsection {The potential conjugate to $Q$}

Quite generally, a global $U(1)$ symmetry in the boundary field theory corresponds to an Abelian gauge symmetry in the bulk.  The charge density and its conjugate chemical potential in the boundary theory can be computed from the corresponding bulk gauge field.  However, in \eqref{F4M2} we did not write down a more general formula in terms of a bulk gauge field as in \eqref{Top2Currents}, but instead we ``solved'' for the electric component of this gauge field in terms of an integration constant $Q$ from the very beginning.  The reason why we did this lies in the intricacies of non-linear consistent truncations:  equation \eqref{F4M2} can probably be generalized to a gauge field with arbitrary components, but one would need to include several other supergravity fields that were consistently set to zero in \eqref{11dMetric}--\eqref{F4M2}.  The reason why in the discussion around equation \eqref{Top2Currents} this was not an issue is that at the linearized level in the gauge field, it is consistent to set these additional supergravity fields to zero.

A generalization of \eqref{F4M2} is still possible without having to turn on other supergravity fields:  one can find the nonlinear generalization of the time-component of the gauge field appearing in \eqref{Top2Currents}.  To find it, one promotes $Q$ to a canonical momentum in the Hamiltonian associated with the $1$-d Lagrangian~\eqref {EffectiveLag}.  Call the canonically conjugate variable $\Phi$. The equation of motion satisfied by $\Phi$ can be found from Hamilton's equation, $\Phi' = {\partial H \over \partial Q}$, which gives
\es{PhiEom}{
\Phi' - \frac {8Q} {r^2} e^{-\frac 12 w-\frac 32\chi} \Bigl(e^{2\eta_1} +
2 e^{2\eta_2}\Bigr)= 0 \,.
}
Plugging $Q$ from eq.~\eqref {PhiEom} in eq.~\eqref {F4M2}, we get
\es {F4M2withPhi} {
 F_4 = -\frac 3L e^{-\frac {21}2\chi} \vol_M - \Phi' \frac {L^3} {e^{2\eta_1} + 2 e^{2\eta_2}} dt \wedge dr \wedge \bigl (e^{2\eta_1} \omega_1 - 2 e^{2\eta_2} \omega_2\bigr)\,.
}
One can explicitly check that this still leads to a consistent truncation.  The equation of motion for $\Phi$ is imposed by the equation of motion for $F_4$.

It is instructive to decompose the form appearing in \eqref{F4M2withPhi} in terms of $\omega$ and $J$,
\es{decomp}{
e^{ 2 \eta_1} \omega_1 - 2 e^{2 \eta_2} \omega_2= \frac {e^{ 2 \eta_1} + 2 e^{2 \eta_2}} {3} \omega
+  \frac {2(e^{ 2 \eta_1} - e^{2 \eta_2})} {3} J  \,,
}
and rewrite eq.~\eqref{F4M2withPhi} as
\es{newffour}{F_4 = -{3\over L} e^{-21 \chi / 2} \vol_M - \Phi' {L^3 \over 3} dt \wedge dr \wedge \omega
	- \Phi' {2 L^3\over 3} {e^{2 \eta_1} - e^{2 \eta_2} \over e^{2 \eta_1} + 2 e^{2 \eta_2} } dt \wedge dr \wedge J \ .}
This shows that $\Phi$ is the Coulomb potential for the topological charge density; for large $r$ it behaves as $-24 Q/r$.  With the boundary conditions described in eq.~\eqref{BoundaryExp}, the last term in $F_4$, which contains $J$, falls off faster than $1/r^2$ and therefore does not correspond to a charge density.

\subsubsection {Thermodynamic quantities}

Thermodynamic quantities in the boundary theory such as the energy density $\epsilon$, entropy density $s$, temperature $T$, $U(1)$ charge density $\rho$, and chemical potential $\mu$ can be calculated from the following formulae
\es{Thermo}{
\epsilon &= -\frac {g_1 e^{-\frac 12 w_0}} {\kappa_4^2 L} \,, \qquad s =
\frac {2 \pi r_h^2} {\kappa_4^2 L^2}\,,
\qquad T = \frac {g'(r_h) e^{-\frac 12 w_h}}
{4\pi}\,,\\
\rho &= \frac {Q} {2 \kappa_4^2} \,, \qquad \mu = \Phi_0 - \Phi_h\,,
}
where the subscript ``$h$'' represents the value of the corresponding field at the horizon, while the subscript ``$0$'' represents the value at the conformal boundary.

There is a simple relation between these quantities,
\es{FirstLaw}{
 \epsilon = \frac 23 (Ts + \mu \rho)\,,
}
which holds in any $(2+1)$-dimensional CFT and can be proven from combining the extensivity relation $\epsilon = T s - p + \mu \rho$ with the tracelessness of the stress-energy tensor $\epsilon = 2 p$.  One can also prove \eqref{FirstLaw} solely from the gravity side by noticing that the ``current''
\es{ConservedCurrent}{
 j \equiv {r^4 \over L^4} e^{{1 \over 2} w} \biggl( {L^2 \over r^2} e^{-w} g  \biggr)' - \frac 18 r^2 \, \frac {e^{\frac 12 w + \frac 32 \chi}}
  {e^{2\eta_1} + 2 e^{2\eta_2}} \, \Phi \Phi'
}
is conserved in the sense that it satisfies $\partial j / \partial r = 0$.  One can check that this current is conserved using the equations of motion following from the effective one-dimensional Lagrangian \eqref{EffectiveLag}.   Evaluated at the horizon, eq.~\eqref{ConservedCurrent} yields
\be
j_h = -2 \kappa_4^2 \,\Phi_h \rho + 2 \kappa_4^2 \,T s\,.
\ee
Evaluated at the conformal boundary, it gives
\be
j_0 = -2 \kappa_4^2 \,\Phi_0 \rho + 3 \kappa_4^2 \, \epsilon\,.
\ee
The equality of the above two relations enforced by the conservation equation yields precisely \eqref{FirstLaw}.

\subsection{Numerics at nonzero temperature}
\label {NUMERICS}

For general values of the parameters, it is unlikely that there are analytic solutions to the equations of motion resulting from the Lagrangian~\eqref {EffectiveLag}. We thus resort to numerical work. We employ a standard shooting technique where we seed the numerical integrator at large $r$, and integrate towards the horizon. The initial conditions are then tuned until a solution that is regular at the horizon is found.

The boundary conditions for solving the equations of motion were described in section~\ref {THERMO}. A series expansion around $r = \infty$ is used to determine the initial conditions for the numerical integration. The first terms in this expansion are given in eq.~\eqref {BoundaryExp}. At fixed $Q$, there are four free parameters, $g_1$, $\chi_6$, $\eta_4$, and $\lambda_2$. One of these parameters can be eliminated by observing that the equations of motion are invariant under the following symmetry transformation:
\be
\label {Symmetry}
g \to \alpha^2 g \,, \qquad r \to \alpha\, r\,, \qquad t \to \alpha^{-1}
t\,, \qquad \vec{x} \to \alpha^{-1} \vec{x}\,, \qquad Q \to \alpha^2 Q\,,
\ee
which can be used to set $g_1= -1$.  The parameters $\chi_6$, $\eta_4$, and $\lambda_2$ can be fixed by imposing the regularity conditions at the horizon, resulting in one solution for every value of $Q$. By varying the dimensionless parameter $Q$ we can probe the boundary field theory at various temperatures, or more precisely, at various values of the dimensionless parameter $T/\mu$.  The $1$-d Lagrangian \eqref{EffectiveLag} is invariant under $Q \to -Q$, so for each solution with a given value of $Q$ one can find another solution by replacing $Q$ by $-Q$.  Without loss of generality, we restrict to the case $Q>0$.

\begin {figure}
\center\includegraphics [width=\textwidth] {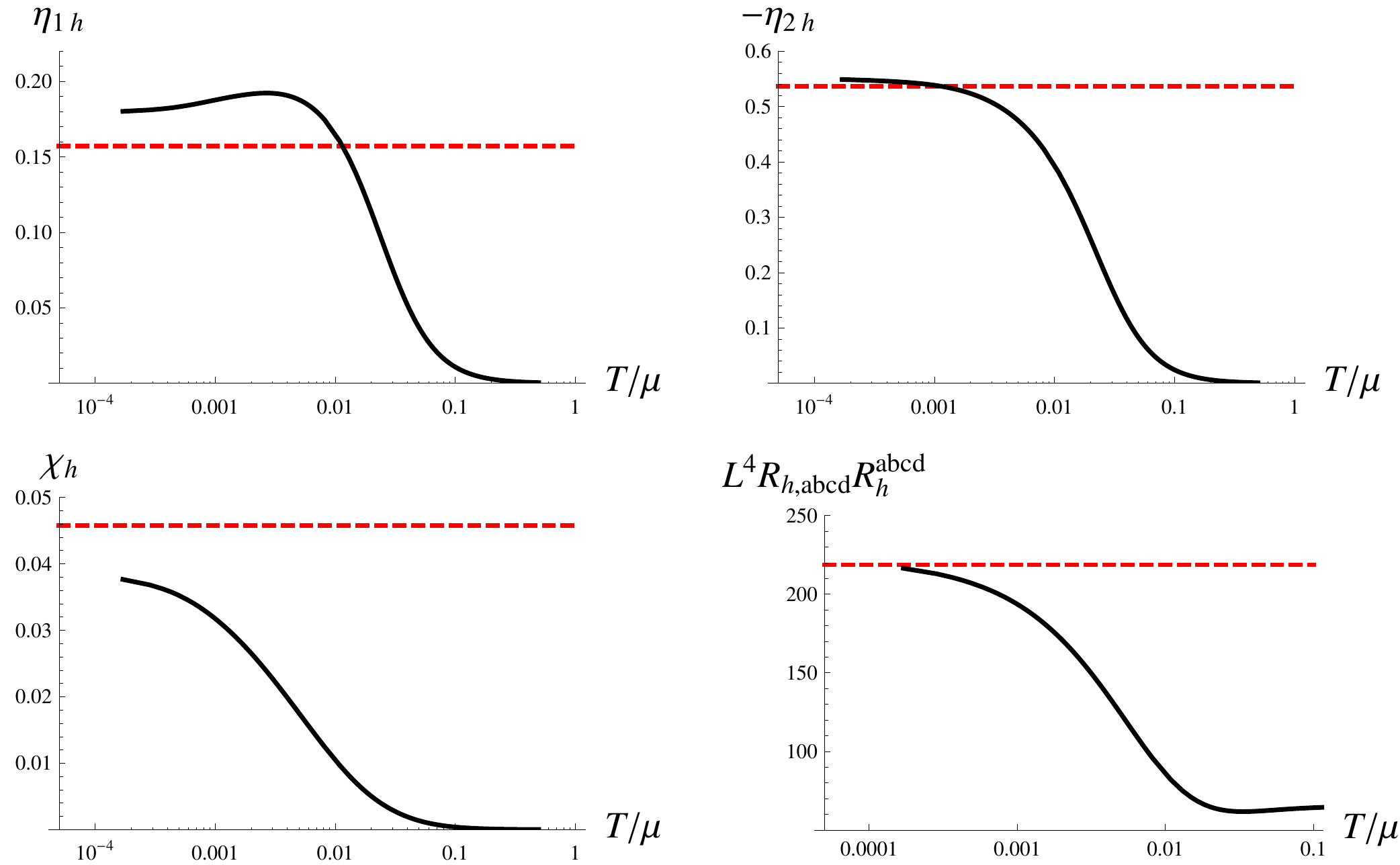}
\caption {The horizon values of the scalars $\eta_1$, $\eta_2$, and $\chi$ and of the squared Riemann tensor as a function of $T/\mu$. The expected zero-temperature values that follow from \eqref{AttractorSoln} are indicated by red dashed lines.  The fact that none of these quantities diverge as $T \to 0$ shows that the supergravity approximation continues to hold down to arbitrarily small temperatures.\label {fig:manyplots}}
\end {figure}

Our numerical results suggest that nothing drastic happens as the temperature approaches zero.  In fact, the scalars $\chi$, $\eta_1$, and $\eta_2$ seem to approach fairly small values at low temperatures:  see figure~\ref{fig:manyplots}.  These values will be computed analytically in the next section.  The bottom right plot in this figure shows that the horizon value of the eleven-dimensional Riemann tensor squared also stays bounded from above as the temperature is decreased.   The lack of divergences means that one can trust the supergravity approximation all the way down to zero temperature.
\begin {figure}
\center\includegraphics [width=0.5\textwidth] {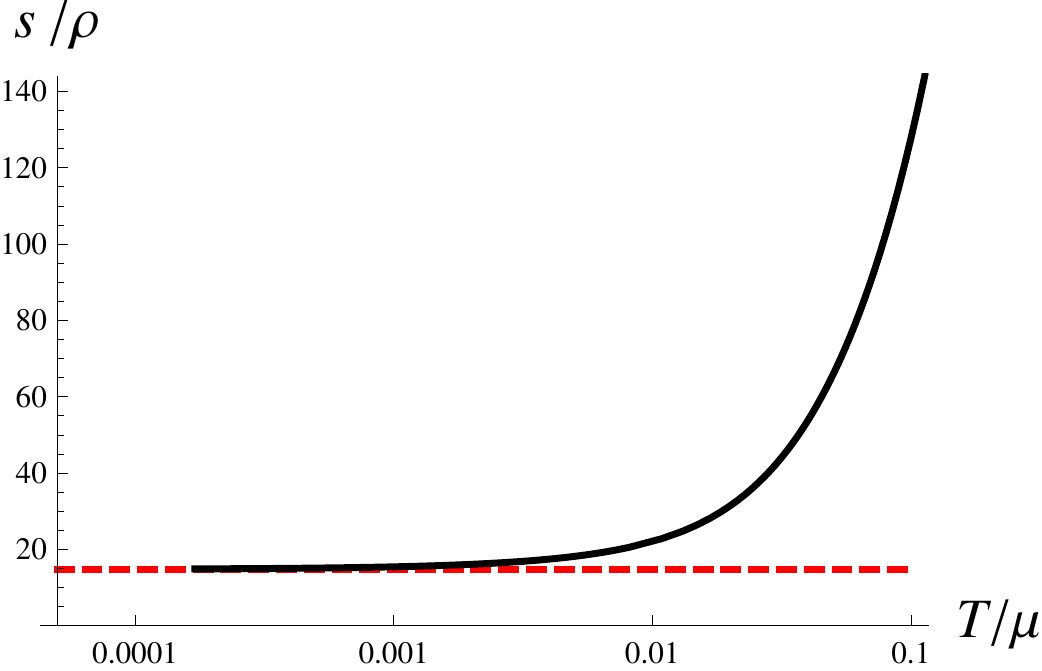}
\caption {The dependence of the ratio of entropy density to charge density on $T/\mu$.  The dashed line indicates the value $s/\rho = 4 \pi / Q \approx 14.75$ expected from the extremal solution of section~\ref{EXTREMAL}.\label {fig:otherplots}}
\end {figure}

\begin {figure}
\center\includegraphics [width=\textwidth] {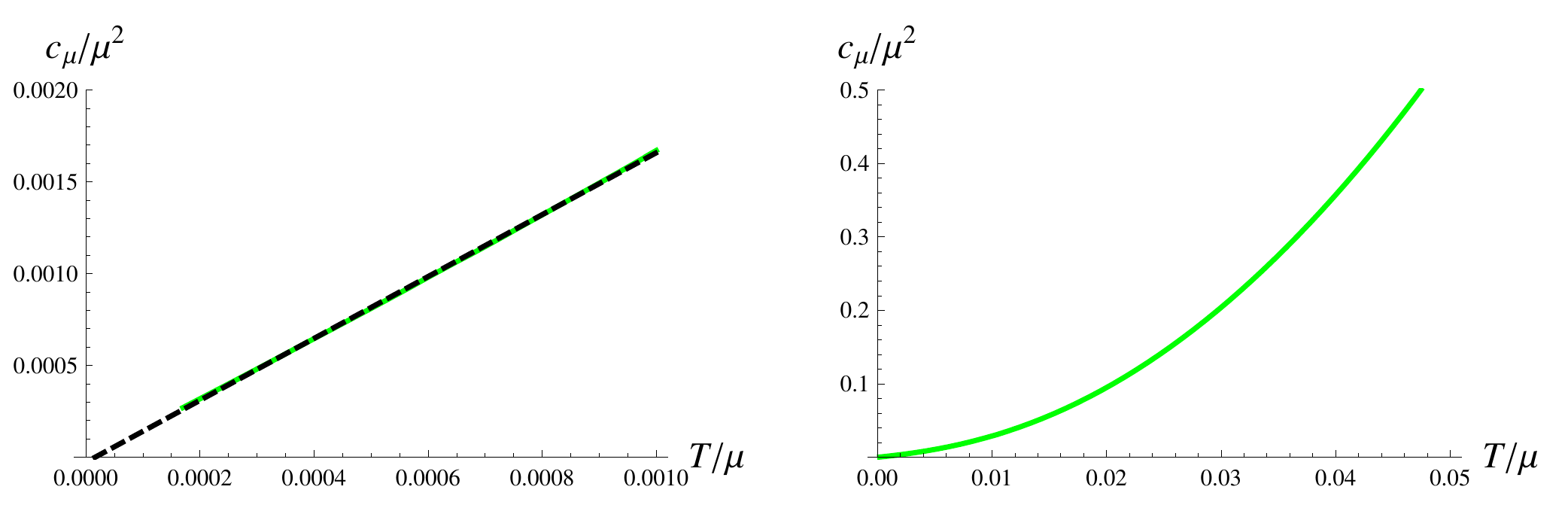}
\caption {The dependence of the specific heat at constant chemical potential on $T/\mu$. The dashed line in the plot on the left is a best fit line, showing the linear behavior of the specific heat at low temperatures.\label {fig:specheat}}
\end {figure}

The thermodynamics of our solutions is similar to that of four-dimensional RNAdS black holes.  For example, for both RNAdS and our backgrounds the entropy density approaches a nonzero value at zero temperature (see figure~\ref{fig:otherplots}).  Similarly, the specific heat at constant chemical potential grows linearly with temperature at low $T$, as can be seen from figure~\ref{fig:specheat}.
In the next section, we will in fact prove that at $T = 0$ the near-horizon four-dimensional geometry is $AdS_2 \times \R^2$, as is also the case for RNAdS\@.

\section{Extremal solutions}
\label{EXTREMAL}

In general, the equations of motion following from \eqref{EffectiveLag} admit black hole solutions with an event horizon at $r = r_h$.  We expect there to exist solutions where the horizon is extremal, which corresponds to having vanishing temperature in the dual field theory.  One of the simplest scenarios is that at extremality $r_h>0$, the functions $\chi$, $\eta_1$, $\eta_2$, and $w$ approach finite values at $r = r_h$, and $g$ behaves as $(r-r_h)^2$ and thus $g'(r_h) = 0$, giving zero temperature by eq.~\eqref{Thermo}.  This scenario describes an extremal horizon which is $AdS_2 \times \R^2 \times \text{squashed } Y^7$, the amount of squashing in the internal space $Y^7$ depending on the values of the scalars at the horizon.  At extremality, $g(r_h) = g'(r_h) = 0$, and the equations of motion following from the Lagrangian \eqref{EffectiveLag} together with the zero-energy constraint \eqref{ZeroEnergy} imply that an $AdS_2 \times \R^2 \times \text{squashed }Y^7$ horizon is possible only if the total potential $V = V_Q + V_s$ (see eq.~\eqref{GotPotentials}) satisfies
\es{ExtHorConditions}{
 V = {\partial V \over \partial \eta_i} = {\partial V \over \partial \chi} = 0 \qquad \text{at $r = r_h$.}
}
These equations are solved by
\es{ScalarsHor}{
 \eta_{1} &= {1 \over 7} \log 3 \,,\qquad
	\eta_{2} = {1 \over 7} \log 3 - \log 2 \,,\qquad
	\chi = {5 \over 14} \log 3 - {1 \over 2} \log 2 \,,
}
as well as
\es{GotQ}{
  Q &= \pm \frac {2^{7 \over 4}} {3^{5 \over 4}} {r_h^2 \over L^2} \approx \pm 0.852\, {r_h^2 \over L^2} \,.
}
We will see shortly that the $AdS_2 \times \R^2 \times \text{squashed } Y^7$ space is in fact an exact solution to the $11$-d supergravity equations of motion for an appropriate choice of the four-form flux $F_4$.

For simplicity in the rest of this section we set $L = r_h = 1$. This can be achieved by using an appropriate choice of units in the bulk to set $L = 1$, and then employing the symmetry~\eqref {Symmetry} to move $r_h$ to $1$.   In this section we will describe three solutions to $11$-d SUGRA:  In section~\ref{ATTRACTOR} we start by describing an analytical solution with seemingly unconventional UV behavior and $AdS_2 \times \R^2 \times \text{squashed }Y^7$ IR asymptotics;  in section~\ref{ADS2SOLN} we recover the analytical solution $AdS_2 \times \R^2 \times \text{squashed }Y^7$ mentioned above as a scaling limit of the solution from section~\ref{ATTRACTOR};  lastly, in section~\ref{NUMERICALZERO} we present a numerical solution with $AdS_4 \times Y^7$ UV asymptotics and $AdS_2 \times \R^2 \times \text{squashed }Y^7$ behavior in the IR\@.

\subsection{A zero-temperature analytical solution}
\label{ATTRACTOR}

The $11$-d SUGRA equations of motion admit the following analytical solution with extremal $AdS_2 \times \R^2 \times \text{squashed }Y^7$ horizon:
\es{MetricF4Attractor}{
 ds^2 &= -{2^3 \over 3^3} e^{-w_0} {(r^4 - 1)^2 \over r^{8 \over 3}} dt^2+ {\sqrt{6} r^{22 \over 3} \over (r^4 - 1)^2} dr^2
  + {2^{7 \over 4} \over 3^{5 \over 4} r^{8 \over 3}} d\vec{x}^2 \\
  {}&+2^{3 \over 2} 3^{1 \over 2} r^{4 \over 3} \left[ ds_{V_1}^2 + {1 \over 2} ds_{V_2}^2
  + {4 \over 3} (d \psi + \sigma_1 + \sigma_2)^2 \right] \,, \\
 F_4 &= -{2^{21 \over 4} \over 3^{11 \over 4}} \vol_M
  - {16 \over 3} \sqrt{2 \over 3} e^{-{1 \over 2} w_0} r^3 dt \wedge dr \wedge (2 \omega_1 - \omega_2) \,.
}
This solution is of the form \eqref{11dMetric}--\eqref{F4M2} given in section~\ref{GEOMETRY} with
\es{AttractorSoln}{
g &= {2^{5 \over 4} (r^4 - 1)^2 \over 3^{7 \over 4} r^{12}}
\,,\qquad w = w_0 - 14 \log r\,, \qquad \Phi = \Phi_h + 4 \sqrt {\frac 23} e^{-\frac 12 w_0} (r^4 - 1)\,,\\
\eta_1 &= {1 \over 7} \log 3 \,,\qquad
\eta_2 = {1 \over 7} \log 3 - \log 2 \,,\qquad
\chi = {1 \over 3} \log {3^{15 \over 14} r^4 \over 2^{3 \over 2}}\,, \qquad Q = \frac {2^{7 \over 4}} {3^{5 \over 4}}  \,,
}
which shows quite explicitly that in the IR the scalars stabilize to the values calculated above in eq.~\eqref{ScalarsHor}, and the charge $Q$ is the same as in \eqref{GotQ}.  There is, of course, another solution to the equations of motion that differs from the one above in the sign of $Q$.

From a field theory perspective, the presence of the $AdS_2$ factor in the IR geometry means that the effective IR field theory can be thought of as a $(0+1)$-dimensional quantum mechanics, which can perhaps arise from a chiral sector of a $(1+1)$-dimensional CFT\@.  The effective dimensions of various operators are related to the IR behavior of supergravity fluctuations around the background \eqref{AttractorSoln}:  a supergravity field dual to an operator ${\cal O}$ of dimension $\Delta_{\rm IR}$ has two linearly independent solutions, one behaving as $(r-1)^{\Delta_{\rm IR}}$ and one as $(r - 1)^{1 - \Delta_{\rm IR}}$ as $r \to 1$.  The coefficient of the first of these two solutions corresponds to a source for ${\cal O}$, while the coefficient of the second one corresponds to an expectation value.

Some of the simplest operators one can study correspond to fluctuations of the fields already present in the consistent truncation \eqref{EffectiveLag}.  It turns out that the linearized equations for the perturbations $(\delta \chi, \delta \eta_1, \delta \eta_2, \delta g, \delta w)$ can be solved exactly.  The solution is
\be
\label {ExtremalPerts}
\begin {aligned}
\delta \eta_1 = c_{\eta_1}\, (r^4 - 1)^\alpha \,, \qquad
\delta \eta_2 &= c_{\eta_2}\, (r^4 - 1)^\alpha \,, \qquad
\delta \chi = c_\chi\, (r^4 - 1)^\alpha \,, \\
\delta w = -21 \,\delta \chi &\,, \qquad \delta g = -\frac {7 \cdot 2^{1/4}}
{3^{3/4} r^{12}} \,c_\chi (r^4 - 1)^{\alpha + 2}\,,
\end {aligned}
\ee
where there are six possible choices for $\alpha$,
\begin {subequations}
\label {AlphaChoices}
\begin {align}
\alpha_1 &= -\frac 12 \pm \frac {\sqrt {69}} 6 \,,\\
\alpha_2 &= -\frac 12 \pm \frac 16 \sqrt {66 - 3 \sqrt {73}}\,,\\
\alpha_3 &= -\frac 12 \pm \frac 16 \sqrt {66 + 3\sqrt {73}} \,.
\end {align}
\end {subequations}
The coefficients $c_i$ are not independent, but are related by the following equations
\be
c_{\eta_1} = -\frac {3 c_\chi} {4} \frac {15 \alpha^2 + 15 \alpha - 28}
{6\alpha^2 + 6\alpha - 7} \,, \qquad
c_{\eta_2} = \frac {3 c_\chi} 8 \frac {126\alpha^4 + 252 \alpha^3 - 177 \alpha^2
-303 \alpha + 140} {6\alpha^2 + 6\alpha - 7}\,.
\ee
These perturbations correspond to three irrelevant operators in the dual quantum mechanics of dimensions
\es{OpDims}{
 \Delta_{{\rm IR},1} = \frac 12 + \frac {\sqrt {69}} 6 \,,
 \qquad \Delta_{{\rm IR},2} = \frac 12 + \frac 16 \sqrt {66 - 3 \sqrt {73}} \,,
 \qquad \Delta_{{\rm IR},3} = \frac 12 + \frac 16 \sqrt {66 + 3 \sqrt{73}}\,.
}
Solutions with different UV behavior for the functions appearing in the $11$-d metric \eqref{11dMetric} (in particular the one with $AdS_4 \times Y^7$ UV asymptotics we will discuss) generate in the IR sources for these operators.  Some fluctuations of $11$-d supergravity not included in the consistent ansatz \eqref{EffectiveLag} are given in Appendix~\ref{FLUCTAPPENDIX}.

\subsection{The IR ``attractor'' as a scaling limit}
\label{ADS2SOLN}

The $AdS_2 \times \R^2 \times \text{squashed }Y^7$ IR asymptotics of the exact solution described in the previous section represent in fact another exact solution to the $11$-d SUGRA equations of motion.  Indeed, the $AdS_2 \times \R^2 \times \text{squashed }Y^7$ ``attractor'' arises as a scaling limit of \eqref{MetricF4Attractor} where one sends $r \to 1 + y \epsilon$ and $t \to t / \epsilon$ and then takes the limit $\epsilon \to 0$.  The background obtained in this limit is $AdS_2 \times \R^2 \times \text{squashed }Y^7$ supported by four-form flux:
\es{AdS2Solution}{
 ds^2 &= -{2^7 \over 3^3} e^{-w_0} y^2  dt^2 + {3^{1 \over 2} \over 2^{7 \over 2}} {1 \over y^2} dy^2 + {2^{7 \over 4} \over 3^{5 \over 4}} d\vec{x}^2 \\
  {}&+ 2^{3 \over 2} 3^{1 \over 2} \left[ ds_{V_1}^2 + {1 \over 2} ds_{V_2}^2
  + {4 \over 3} (d \psi + \sigma_1 + \sigma_2)^2 \right] \,, \\
 F_4 &= -{2^{21 \over 4} \over 3^{11 \over 4}} e^{-{1 \over 2} w_0} dt \wedge dx^1 \wedge dx^2 \wedge dy - {16 \over 3} \sqrt{2 \over 3} e^{-{1 \over 2} w_0} dt \wedge dy \wedge (2 \omega_1 - \omega_2) \,.
}
Note that this solution is not of the form \eqref{11dMetric}--\eqref{F4M2} because the coefficient of $d\vec{x}^2$ in \eqref{MetricM} cannot be set to a constant.  Perturbations around this solution can be computed directly from perturbing the $11$-d background \eqref{AdS2Solution}, or can be obtained by taking the scaling limit of perturbations around the background \eqref{MetricF4Attractor} such as \eqref{ExtremalPerts}.

\subsection {A numerical solution with $AdS_4 \times Y^7$ asymptotics}
\label{NUMERICALZERO}

Three of the six linearly independent perturbations described in section~\ref{ATTRACTOR}, namely the ones corresponding to sources for the operators of dimensions \eqref{OpDims}, are well-behaved at the horizon. These three integration constants allow us, at least at the linearized level, to adjust to zero the asymptotic values of the scalars at large $r$ so that our solutions asymptote to $AdS_4 \times Y^7$.  Of course, there is no guarantee that the same holds true for the exact equations, but we can check numerically that this is indeed the case. As before, we use a standard shooting technique, this time seeding the numerical integrator very close to the horizon. We use the linearized perturbations as a seed, and tweak the coefficients of the three linearly independent perturbations until we find a solution that obeys the desired boundary conditions at large $r$.

\begin {figure}
\center\includegraphics [width=0.9\textwidth] {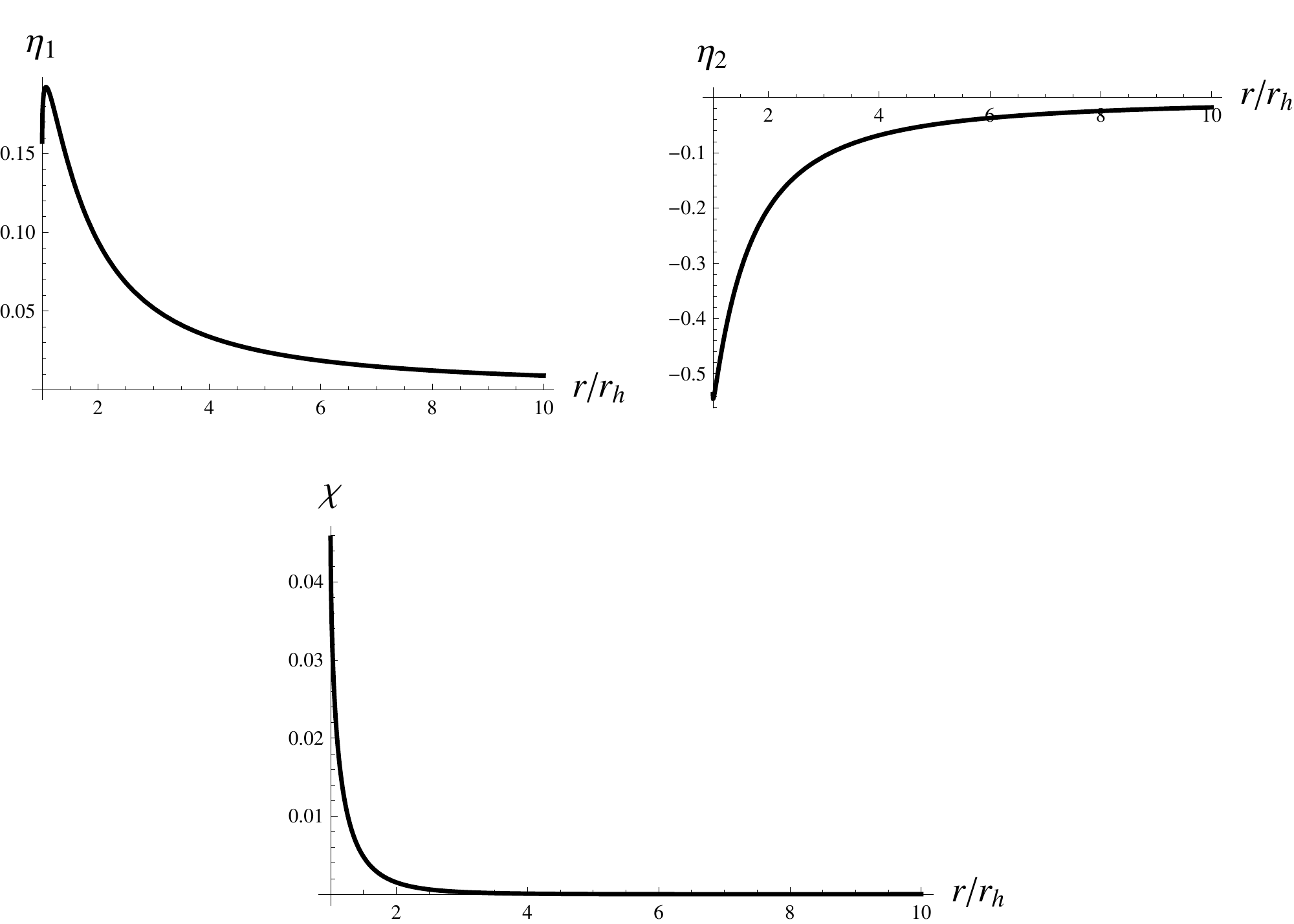}
\caption {The dependence of the scalars $\eta_1$, $\eta_2$, and $\chi$ on the radial variable $r$ at zero temperature.  We see that the scalars tend to zero at the boundary since our solution asymptotes to $AdS_4 \times Y^7$ in the UV\@.\label {fig:scalarszeroT}}
\end {figure}

Plots showing the behavior of the scalars as a function of radial coordinate are given in figure~\ref {fig:scalarszeroT}. We thus see that there exists an extremal black hole solution that interpolates between the attractor solution of the previous section in the IR and $AdS_4 \times Y^7$ in the UV\@.  As a consistency check, we verified that our zero-temperature numerics are consistent with $s/\rho = 4 \pi / Q \approx 14.75$ and $R_{h, abcd} R_h^{abcd} = 656/3 \approx 218.67$, which can be calculated directly from the attractor solution \eqref{AttractorSoln}, as these quantities are insensitive to the UV asymptotics.  These values are also consistent with the finite-temperature numerics that we discussed in section~\ref{NUMERICS};  see figures~\ref{fig:manyplots} and~\ref{fig:otherplots}.

\section{The potential for probe M2-branes}
\label{M2FORCE}

There are two types of M2-branes present in our construction: the M2-branes filling the $(t,x^1,x^2)$ directions, which are responsible for generating the asymptotic $AdS_4\times Y^7$ space, and M2-branes wrapped over a two-cycle in the internal space which are responsible for the topological charge of the membrane solution.
We will henceforth refer to the former type of branes as space-time filling, and to the latter as wrapped.

One might wonder whether there is an instability where any of these branes tunnel out to infinity \cite{Hartnoll:2009ns,Yamada:2007gb, Yamada:2008em}.  We investigate this question by computing the potential for a probe brane as a function of the AdS radial variable $r$. The action for a probe brane is
\es{M2Action}{
S_{M2} = -\tau_{M2} \int d^3 x\, \sqrt{-G} \pm \tau_{M2} \int A_3 \,,
}
where $\tau_{M2}$ is the M2-brane tension,
\es {M2Tension}{
\tau_{M2} = \frac {2\pi} {(2\pi \ell_p)^3}\,,
}
and $A_3$ is the three-form gauge potential for $F_4 = dA_3$. We are primarily interested in the sign such that the interaction with $A_3$ is repulsive, i.e.~when the M2-brane has the same charge as the stack that creates our background. Then the force on the brane vanishes in $AdS_4\times Y^7$. The opposite sign corresponds to a probe anti M2-brane, for which the force is attractive at infinity.

For static embeddings, one can define a potential $V$ for the probe branes through
\es {M2PotDef}{
S_{M2} = -\int V \, dt \,.
}
Our backgrounds are metastable if the potential is smaller at some $r>r_h$ than at the horizon.\footnote{SSP thanks A.~Yarom for a discussion on this issue.}

\subsection{Probe space-time filling M2-branes}
\label{M2FILLING}

\begin {figure}
\center\includegraphics [width=0.8\textwidth] {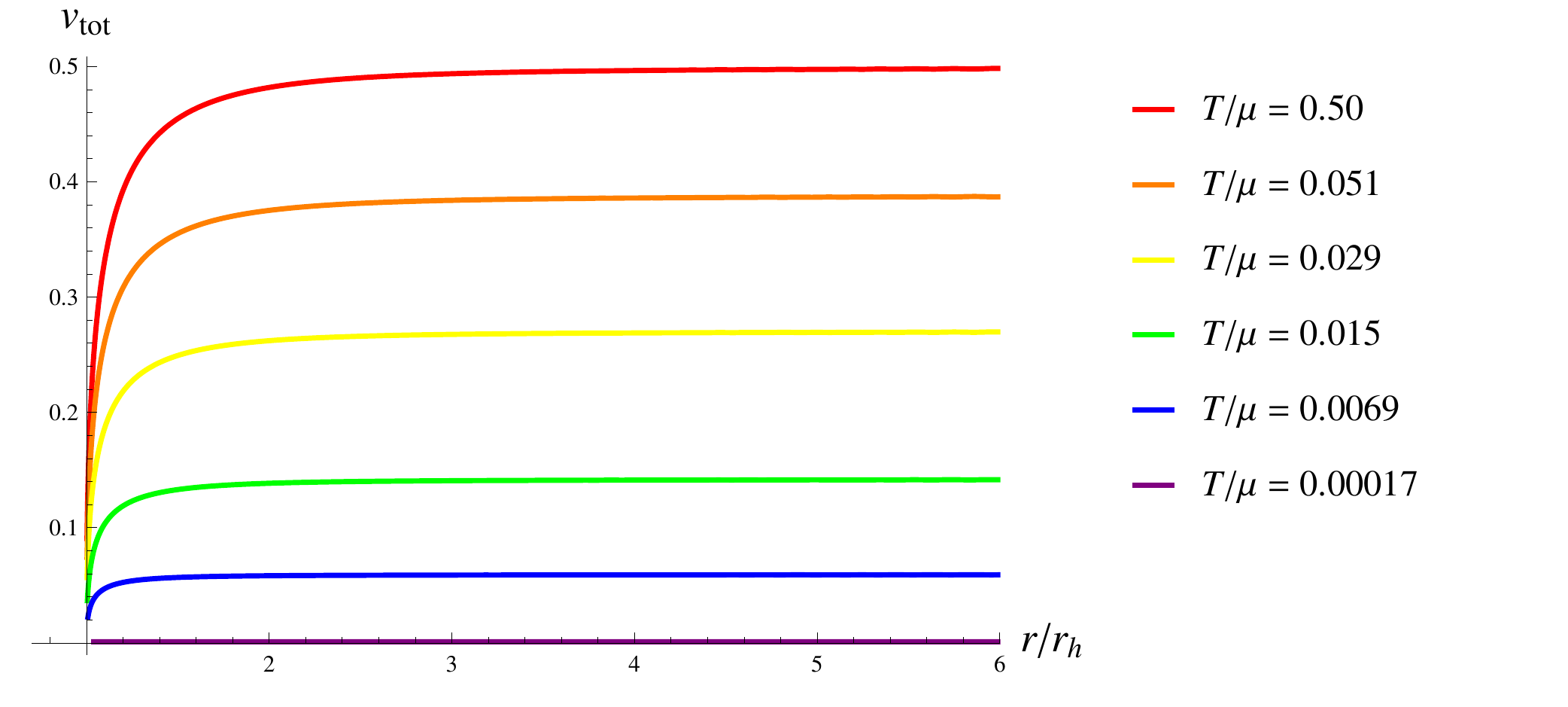}
\caption {The potential energy per unit area of a probe space-time filling M2-brane as a function of the AdS radial coordinate $r$, at various temperatures. We worked in a gauge where the horizon value of the potential vanishes.\label {fig:fillingbraneplot}}
\end {figure}

Since the volume of these branes is infinite, we will look at their potential energy per unit area. We thus write
\newcommand {\potfill} {v}
\es {FillingM2Action}{
S_{M2} = -\int dt \,d^2x \, \bigl[ \potfill_g (r) + \potfill_e (r)\bigr]\,,
}
where $\potfill_g (r)$ and $\potfill_e (r)$ come from the first and second terms in \eqref{M2Action}, respectively. It is straightforward to calculate these contributions using eq.~\eqref {11dMetric}. We have
\es {FillingPotentials} {
\potfill_g (r) = \tau_{M2} r^2 \sqrt {g} e^{-\frac 12 w - \frac {21} 4 \chi} \,,\qquad \potfill'_e (r) = \mp 3 \tau_{M2} r^2 e^{-\frac 12 w - \frac {21} 2 \chi}\,,
}
and we can choose, for example, $\potfill_e(r_h) = 0$.  Here and in the rest of this section we set $L = 1$.  The minus sign in $\potfill_e(r)$ corresponds to probe branes, while the plus sign corresponds to probe anti-branes.  The probe anti-branes are always attracted towards the horizon, so we will only focus on the probe branes.  In figure~\ref {fig:fillingbraneplot}, we have plotted the potential $\potfill_{\rm tot}(r) \equiv \potfill_g(r) + \potfill_e(r)$ at various temperatures, as a function of $r$. We see that the potential never dips below the horizon value, so the background is stable with respect to tunneling of space-time filling M2-branes.

One can also evaluate the potential $\potfill_{\rm tot} (r)$ for the space-time filling branes on the analytical extremal solution~\eqref {AttractorSoln}.  In this case, the potential vanishes identically.  From the plots in figure~\ref {fig:fillingbraneplot}, it looks like the space-time filling M2-brane potential also vanishes identically in the extremal limit of the solution of section~\ref{NUMERICALZERO} that asymptotes to $AdS_4\times Y^7$, so one might wonder whether this result is insensitive to the UV asymptotics of the solution.  Indeed, one can prove this result starting with the observation that the force per unit area, $f_{\rm tot} \equiv -\potfill_{\rm tot}'$, satisfies the following first order differential equation:
\es{ForceEq}{
f_{\rm tot}' + \left[{3 e^{-{21 \over 4} \chi} \over \sqrt{g}}
 + {g' \over 2 g} + {21 \chi' \over 4} \right] f_{\rm tot}= 0 \,.
}
Since this equation is linear, its solutions depend on one integration constant that acts as a multiplicative factor.  Near the extremal horizon, \eqref{AttractorSoln} and \eqref{ExtremalPerts} give
\es{CoeffHor}{
{3 e^{-{21 \over 4} \chi} \over \sqrt{g}}
 + {g' \over 2 g} + {21 \chi' \over 4} =
 {2 \over r -1} + \text{subleading} \,,
}
so
\es{GotfHor}{
f_{\rm tot} = c \left[(r-1)^2 + \text{subleading} \right] \,.
}
The subleading terms in the above two equations are sensitive to the UV asymptotics, but the leading term is not.  Since $\potfill_{\rm tot}$ vanishes identically when evaluated on the leading behavior \eqref{AttractorSoln}, it must be that $c = 0$.  Therefore, the potential for the space-time filling M2-branes is exactly flat for any solution that connects to the solution \eqref{AttractorSoln} in the IR\@.  It is worth noting that even though the exact $AdS_2 \times \R^2 \times \text{squashed }Y^7$ solution \eqref{AdS2Solution} cannot be written in the gauge \eqref{11dMetric}--\eqref{MetricM}, one can also show that the potential for space-time filling branes is exactly flat in this case too.  The flatness of the potential follows from taking a scaling limit of the exact solution \eqref{MetricF4Attractor} as explained at the beginning of section~\ref{ADS2SOLN}.

The existence of a flat potential for the space-time filling branes is reminiscent of supersymmetric solutions, so one might wonder whether our background preserves any supersymmetry.  In a supersymmetric background, the gravitino variation
\es {KillingFermions} {
  \delta_\mu \epsilon \equiv \nabla_\mu \epsilon + \frac 1{12} \biggl (\frac 1 {4!} F_{\nu\rho\lambda\sigma} \Gamma_\mu \Gamma^{\nu\rho\lambda\sigma} - \frac 12 F_{\mu\nu\rho\lambda} \Gamma^{\nu\rho\lambda}\biggr) \epsilon
}
vanishes identically.  A necessary condition for this to happen is that
\es {KillingCommutator} {
  [\delta_\mu, \delta_\nu] \epsilon = 0\,,
}
which is a linear system of algebraic equations.  One can check that this system has no non-trivial solutions for both the backgrounds~\eqref {MetricF4Attractor} and~\eqref {AdS2Solution}.

\subsection{Probe wrapped M2-branes}
\label{M2GENERALWRAPPED}

Let us consider a static M2-brane embedding where the brane wraps a topologically non-trivial two-dimensional cycle ${\cal C}$ in the internal space and sits at some fixed values of $r$ and $\vec{x}$.  By the internal space we mean the squashed version $\tilde Y^7$ of $Y^7$ appearing in \eqref{11dMetric} with the metric
\es{tildeYMetric}{
ds_{\tilde Y}^2 =  e^\chi \left[
e^{\eta_1} ds_{V_1}^2 + e^{\eta_2} ds_{V_2}^2
+ e^{-4 \eta_1 - 2 \eta_2} (d \psi + \sigma_1 + \sigma_2)^2  \right] \,.
}
For a fixed value of $r$ at which the scalars $\chi$, $\eta_1$, and $\eta_2$ don't diverge, the topology of $\tilde Y^7$ is the same as that of $Y^7$, and there is a one-to-one correspondence between surfaces in $Y^7$ and surfaces in $\tilde Y^7$.  So when we say that a brane wraps a cycle ${\cal C}$ in $\tilde Y^7$, we might as well be thinking about the corresponding cycle in $Y^7$, and indeed we will not be careful about this distinction in the rest of this section unless there is potential for confusion.

The probe-brane action \eqref{M2Action} takes the form
\es{PotentialFromAction}{
S_{M2} = - \int dt \left[ V_g^{\cal C}(r) + V_e^{\cal C}(r) \right] \,,
}
where $V_g^{\cal C}(r)$ and $V_e^{\cal C}(r)$ come from the first and second terms in \eqref{M2Action}, respectively.  We will call $V_g^{\cal C}(r)$ the gravitational potential and $V_e^{\cal C}(r)$ the electrostatic potential for such a brane.  Stable brane wrappings are of course those that minimize the total potential $V_{\rm tot}^{\cal C}(r) \equiv V_g^{\cal C}(r) + V_e^{\cal C}(r)$.

A simple way to construct non-trivial two-cycles in $\tilde Y^7$ is to start with a two-cycle in the base $V = V_1 \times V_2$, and lift it to $\tilde Y^7$.  However, not every two-cycle in the base can be lifted to a two-cycle in the total space. The reason for this restriction is that when lifting a two-cycle, one needs to specify what the fiber angle should be at all points on that cycle, and such an assignment may not be consistent because of topological reasons.  We include a more technical discussion of these issues in appendix~\ref {APPENDIXCYCLES}.  The upshot is that any (well-defined) two-cycle $\mathcal C$ in $\tilde Y^7$ satisfies
\es{LiftableConditionSimp}{
\int_{\mathcal C} J = 0 \,,
}
where $J = \omega_1 + \omega_2$, as in \eqref{KahlerForm}.   One way to understand this condition is to note that $J$ is a closed form in $Y^7$ because it obeys $d e_\psi = 2 J$, where $e_\psi \equiv d \psi + \sigma_1 + \sigma_2$ is a globally-defined one-form on $Y^7$.\footnote{Note that $\sigma_1 + \sigma_2$ by itself is not a globally defined one-form, so the condition \eqref{LiftableConditionSimp} does not hold for two-cycles in $V_1 \times V_2$.}

Using \eqref{newffour} and the fact that $\int_{\cal C} J = 0$, one can write the electrostatic potential $V_e^{\cal C}(r)$ as
\es{SecondTerm}{
V_e^{\cal C}(r) = \mp {1 \over 3} \tau_{M2} \Phi \int_{\cal C} (\omega_1 - 2 \omega_2)
= \pm \tau_{M2} \Phi  \int_{\cal C} \omega_2 \,.
}
This term depends only on the homology class of ${\cal C}$ in $H_2(Y^7; \Z)$, so in order to find the stable wrappings for a given homology class $H_2(Y^7; \Z)$ one has to minimize only the gravitational potential $V_g^{\cal C}(r)$.  Two questions arise:
\begin{enumerate}[(I)]
\item \label{FirstQuestion} For a static M2-brane embedding at fixed $r$, what cycles ${\cal C}$ are stable in the sense that they minimize $V_{\rm tot}^{\cal C}(r)$, at least compared to neighboring cycles?  Since $V_e^{\cal C}(r)$ is topological and $V_g^{\cal C}(r)$ is proportional to the volume (or more correctly, area) of ${\cal C}$ computed using the induced metric from $\tilde Y^7$, this problem reduces to finding the minimal volume cycles of $\tilde Y^7$.
\item \label{SecondQuestion} How does the minimal value of $V_{\rm tot}^{\cal C}(r)$ from \eqref{FirstQuestion} depend on $r$?  Are the branes repelled from the black hole horizon, or do they tend to fall into the black hole?
\end{enumerate}

The first question is interesting in its own right, but may be hard to answer in general, especially since there are Sasaki-Einstein manifolds such as the spaces ${\cal P}_n$ described in section~\ref{EXAMPLES} for which the metric is not known explicitly.  We will therefore content ourselves with finding a lower bound on the volumes of the cycles ${\cal C}$ of $\tilde Y^7$ in the cases where $Y^7$ is a regular Sasaki-Einstein manifold.  Such a bound can be found by using calibrations, as we discuss in appendix~\ref{APPENDIXCALIB}.  This bound is
\es{VolumeIneqSimp}{
\Vol({\cal C}) \geq e^{\chi} \left( e^{\eta_1} + e^{\eta_2} \right) \left| \int_{\mathcal C} \omega_2 \right| \,.
}
For an arbitrary homology class in $H_2(\tilde Y^7; \Z)$, this inequality may not be saturated by any embedded surfaces in that class.  However, as we now explain, the bound \eqref{VolumeIneqSimp} is restrictive enough to show that wrapped M2-branes do not condense.

\begin {figure}
\center\includegraphics [width=0.8\textwidth] {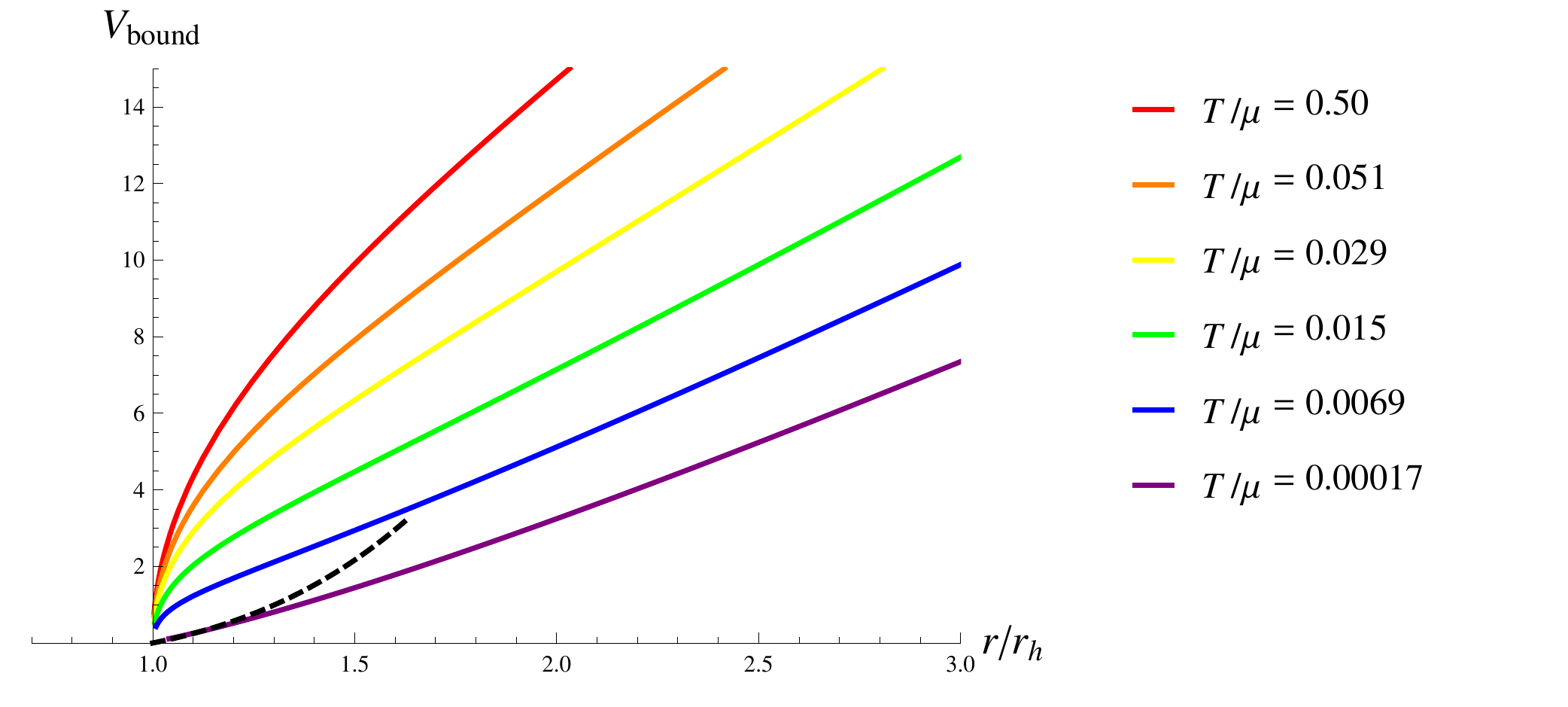}
\caption {The bound~\eqref {VtotBound} for the potential for probe M2-branes wrapping a two-cycle ${\cal C}$, expressed as multiples of $\tau_{M2} \bigl\lvert \int_{\mathcal C} \omega_2 \bigr\rvert$ and normalized so that it vanishes at the horizon.  Each solid curve corresponds to a different temperature. The dashed line represents the analytic approximation~\eqref {M2VgeM111ext}, valid close to the extremal horizon.  This bound is saturated for the cycles~\eqref {CM111} and \eqref{C2Q111} in $M^{1,1,1}$ and $Q^{1,1,1}$, respectively.\label {fig:wrappedbraneplot}}
\end {figure}

Equation \eqref{VolumeIneqSimp} can be used to find a lower bound on the gravitational potential for a wrapped M2-brane:
\es{M2VgBound}{
V_{g}^{\cal C}(r) &= 4 \tau_{M2} e^{-{7 \over 4} \chi} \sqrt{g} e^{-{w \over 2}} \Vol({\cal C})
\geq 4 \tau_{M2} e^{-{3 \over 4} \chi} \sqrt{g} e^{-{w \over 2}} \left( e^{\eta_1} + e^{\eta_2} \right) \left| \int_{\mathcal C} \omega_2 \right| \,.
}
Combining this equation with the expression for the electrostatic potential \eqref{SecondTerm}, we find that the total potential satisfies
\es{VtotBound}{
V_{\rm tot}^{\cal C} \geq V_{\rm bound}^{\cal C} \,, \qquad
V_{\rm bound}^{\cal C} \equiv \tau_{M2} \left[ 4 e^{-{3 \over 4} \chi} \sqrt{g} e^{-{w \over 2}} \left( e^{\eta_1} + e^{\eta_2} \right)
 - \Phi \right]  \left| \int_{\mathcal C} \omega_2 \right| \,.
}
From eq.~\eqref {M2VgBound} it also follows that at the horizon $V_{\text{tot}}^{\mathcal {C}} (r_h) = V_{\text{bound}}^{\mathcal {C}} (r_h)$, because $V_g^{\cal C}(r_h) = 0$.  From figure~\ref {fig:wrappedbraneplot}, we see that in a gauge where $\Phi$ vanishes at the horizon, $V_{\text {bound}}^{\mathcal {C}} (r) > 0$ for all $r > r_h$, implying that
\es {WrappedStable}{
V_{\text {tot}}^{\mathcal {C}} (r) > V_{\text {tot}}^{\mathcal C} (r_h)\,, \qquad \text {for $r > r_h$.}
}
This inequality means that the wrapped M2-branes do not condense.

One can check analytically that the wrapped M2-branes are attracted by the horizon at extremality by evaluating the lower bound in \eqref{M2VgBound} and the electrostatic potential \eqref{SecondTerm} on the exact solution \eqref{AttractorSoln}.  The result is
\es{M2VgeM111ext}{
V_{g,\text{bound}}^{\cal C,\text{extremal}}(r) &= 4 \tau_{M2} e^{-{w_0 \over 2}} (r^4 - 1) \left \lvert \int_{\mathcal C} \omega_2 \right\rvert \,,\\
V_{e}^{\cal C,\text{extremal}}(r) &= \pm 4 \sqrt{2\over 3} \tau_{M2} e^{-{w_0 \over 2}} (r^4 - 1) \left ( \int_{\mathcal C} \omega_2 \right)\,.
}
One can see that the gravitational force (which is always inwards) is larger in magnitude than the electrostatic one by at least a factor of $\sqrt{3 \over 2}$, so all these branes tend to fall into the black hole horizon at extremality.  (See figure~\ref {fig:wrappedbraneplot} for a comparison between the analytic formulae~\eqref {M2VgeM111ext} and the numerical results.)  By taking a scaling limit of the exact solution \eqref{AttractorSoln} one can show that these wrapped branes are also always attracted by the extremal horizon in the case of the $AdS_2 \times \R^2 \times \text{squashed }Y^7$ solution \eqref{AdS2Solution}.

\subsubsection{Example 1:  Probe branes wrapping a two-cycle in $M^{1, 1, 1}$}
\label{M111}

The manifold $M^{1, 1, 1}$ is a $U(1)$ fiber bundle over $\CP^2 \times S^2$.  It can be parameterized by seven angles:  $\mu$, $\theta_1$, $\phi_1$, and $\psi_1$ parameterizing $\CP^2$, $\theta_2$ and $\phi_2$ parameterizing $S^2$, and $\psi$ parameterizing the fiber.  In another description, $M^{1, 1, 1}$ is a $U(1)$ quotient of $S^5 \times S^3$.  One can parameterize $S^5$ by three complex coordinates $u^i$, $i = 1, 2, 3$, with $\abs{u^1}^2 + \abs{u^2}^2 + \abs{u^3}^2 = \text{const}$ and $S^3$ by two complex coordinates $v^j$, $j = 1, 2$, satisfying $\abs{v^1}^2 + \abs{v^2}^2 = \text{const}$.  The $U(1)$ quotient acts by identifying $u^i \sim e^{2 i \delta} u^i$ and $v^j \sim e^{-3 i \delta} v^j$.  An explicit Einstein metric on $M^{1, 1, 1}$ as well as more details on this space such as topological properties or the relation between the $(u^i, v^j)$ coordinates and the angular ones can be found in Appendix~\ref{M111METRIC}.

As mentioned above, we want to find the two-cycles of $M^{1, 1, 1}$ (or of the squashed variant thereof $\tilde M^{1, 1, 1}$ as in \eqref{tildeYMetric}) that are local volume minimizers in their homology class. The second homology of $M^{1, 1,1}$ is $H_2(M^{1, 1,1}; \Z) \cong \Z$, so there is only one generator class for it.  A minimal volume cycle representing the generator of the second homology of $M^{1, 1, 1}$ is
\es{CM111}{
{\cal C} : \left\{
\begin{aligned}
\theta_ 1 &= 2 \arctan t^2 & \mu &= {\pi \over 2} \\
\theta_2 &= 2 \arctan t^3 & \psi_1 &= \text{const.}\\
\phi_1 &= 2 \phi & \psi &= \text{const.} \\
\phi_2 &=-3 \phi \,.
\end{aligned}
\right.
\quad
\Longleftrightarrow
\quad
\left\{
\begin{aligned}
\left( {u^1 \over u^2} \right)^2 &= \left({\bar v_1 \over \bar v_2} \right)^3 \\
u^3 &= 0 \,.
\end{aligned}
\right.
}
In order to cover ${\cal C}$ only once, the ranges of $t$ and $\phi$ should be taken to be $t \geq 0 $ and $0 \leq \phi \leq 2 \pi$.\footnote{Similar cycles have been considered in five-dimensional Sasaki-Einstein manifolds.  See for example \cite{Arean:2004mm}.} This cycle is well-defined as it satisfies eq.~\eqref {LiftableConditionSimp} (see also the discussion at the end of Appendix~\ref{APPENDIXCYCLES}), and has minimal volume since it saturates the bound~\eqref {VolumeIneqSimp}, as can be checked by direct computation.

Using the explicit metric on $M^{1, 1,1}$ given in Appendix~\ref{M111METRIC} and the explicit parameterization of the cycle \eqref{CM111}, one obtains the following gravitational potential:
\es{M2VgM111}{
V_{g}^{\cal C}(r) &= 6 \pi \tau_{M2} e^{-{3 \over 4} \chi} \sqrt{g} e^{-{w \over 2}} (e^{\eta_1} + e^{\eta_2}) \,.
}
Similarly, one can use \eqref{M2Action} to find the electrostatic potential
\es{M2VeM111}{
V_{e}^{\cal C}(r) &= \mp {3 \pi \over 2} \tau_{M2} \Phi \,. 
}
The potential for these branes saturates the bound~\eqref {VtotBound}, as a consequence of the fact that the cycle they wrap saturates~\eqref {VolumeIneqSimp}.

\subsubsection{Example 2:  Probe branes wrapping a two-cycle in $Q^{1, 1, 1}$}
\label{Q111}

The manifold $Q^{1, 1, 1}$ can be described as a $U(1)$ fibration over $S^2 \times S^2 \times S^2$, so it can be parameterized in terms of three sets of angles $(\theta_a, \phi_a)$, $a = 1, 2, 3$, each set parameterizing one of the spheres, and a fiber angle $\psi$.  Another description of $Q^{1, 1, 1}$ is as a $U(1)^2$ quotient of $S^3 \times S^3 \times S^3$:  One can parameterize the $S^3$'s by three sets of two complex coordinates, $a^i$, $b^j$, $c^k$, with $i, j, k = 1, 2$, satisfying $\abs{a^1}^2 + \abs{a^2}^2 = \abs{b^1}^2 + \abs{b^2}^2 = \abs{c^1}^2 + \abs{c^2}^2 = \text{const}$, and take a quotient by a $U(1)$ that acts by $a^i \sim e^{i \delta} a^i$, $b^j \sim e^{-i \delta} b^j$, $c^k \sim c^k$, and by another $U(1)$ that acts by $a^i \sim e^{i \delta} a^i$, $b^j \sim b^j$, $c^k \sim e^{-i \delta} c^k$.  More details about $Q^{1, 1, 1}$ including an explicit Einstein metric, the relation between the complex coordinates and the angles, and some information about its topology are given in Appendix~\ref{Q111METRIC}.

The second homology of $Q^{1, 1, 1}$ is $H_2(Q^{1, 1, 1}; \Z) \cong \Z^2$.  Its generators can be represented by the following minimal volume cycles
\es{C1Q111}{
{\cal C}_1 : \left\{
\begin{aligned}
\theta_ 1 &= \theta_2 \\
\theta_3 &= \text{const.} \\
\phi_1 &= - \phi_2 \\
\phi_3 &=  \text{const.}
\end{aligned}
\right.
\quad
\Longleftrightarrow
\quad
\left\{
\begin{aligned}
{a^1 \over a^2} &= {\bar b_1 \over \bar b_2} \\
c^1 &= \text{const.} \\
c^2 &= \text{const.}
\end{aligned}
\right.
}

\es{C2Q111}{
{\cal C}_2 : \left\{
\begin{aligned}
\theta_ 1 &= \theta_3 \\
\theta_2 &= \text{const.} \\
\phi_1 &= - \phi_3 \\
\phi_2 &=  \text{const.}
\end{aligned}
\right.
\quad
\Longleftrightarrow
\quad
\left\{
\begin{aligned}
{a^1 \over a^2} &= {\bar c_1 \over \bar c_2} \\
b^1 &= \text{const.} \\
b^2 &= \text{const.}
\end{aligned}
\right.
}
It is straightforward to compute the gravitational and electrostatic potentials for probe M2-branes wrapping these cycles:
\es{Q111Potentials}{
V_g^{{\cal C}_1}(r) &= 4 \pi \tau_{M2} e^{-{3 \over 4} \chi} \sqrt{g} e^{-{w \over 2}} e^{\eta_1} \,,
\qquad V_e^{{\cal C}_1}(r) = 0 \,, \\
V_g^{{\cal C}_2}(r) &= 2 \pi \tau_{M2} e^{-{3 \over 4} \chi} \sqrt{g} e^{-{w \over 2}} (e^{\eta_1} + e^{\eta_2}) \,, \qquad
V_{e}^{{\cal C}_2}(r) = \mp {\pi \over 2} \tau_{M2} \Phi \,.
}
The cycle ${\cal C}_2$ saturates the bounds \eqref{VolumeIneqSimp} and \eqref{VtotBound}, while ${\cal C}_1$ does not.

\section{Field theory interpretation}
\label{FIELDTHEORY}

Let us discuss a dual 3-d gauge theory interpretation of our brane solutions carrying topological charges. The solutions are asymptotic to $AdS_4\times Y^7$, with the Sasaki-Einstein space $Y^7$ having $b_2>0$. The classic examples of such backgrounds known since the 80's are $AdS_4\times M^{1,1,1}$,  $AdS_4\times Q^{1,1,1}$, and $AdS_4\times Q^{2,2,2}$. The search for the 3-d ${\cal N}=2$ superconformal field theories dual to them began in the late 90's; see, for example, \cite{Fabbri:1999hw}. Following the major progress on formulating the world volume theories of coincident M2-branes
\cite{Bagger:2006sk, Bagger:2007jr, Bagger:2007vi,Gustavsson:2007vu,Aharony:2008ug}, a recent wave of research has produced compelling proposals for the Chern-Simons (C-S) quiver gauge theories dual to these M-theory backgrounds \cite{Martelli:2008si,Hanany:2008cd,Franco:2008um,Franco:2009sp,Davey:2009sr}. Interestingly, all these proposals involve $U(N)^{2+ b_2}$ gauge theories with a certain set of Chern-Simons levels $k_1, k_2, \ldots, k_{2+b_2}$ that add up to zero.

The discussion of the Abelian $U(1)^{2+b_2}$ subgroup of the gauge group requires special care. None of the matter fields are charged under the diagonal $U(1)$ corresponding to the gauge field ${\cal A}_+\sim \sum_{j=1}^{2+ b_2} {\cal A}_j$. The existence of magnetic monopole configurations for this diagonal $U(1)$ means that another combination of the $U(1)$'s ${\cal A}_b \sim \sum_{j=1}^{2+ b_2} k_j {\cal A}_j$ gets gauge fixed to a discrete subgroup. The remaining $b_2$ gauge fields
\es{orthogonal}{ {\cal A}_{\vec m}\sim \sum_{j=1}^{2+ b_2} m_j {\cal A}_j
}
may be chosen to be orthogonal to each other; they are orthogonal to ${\cal A}_+$ and ${\cal A}_b$ due to the conditions $\vec m \cdot \vec k=\vec m \cdot \vec I=0$, where $\vec I = (1, 1, \ldots, 1)$. The gauge fields ${\cal A}_{\vec m}$ have C-S terms and are coupled to massless charged matter. For each of them
one can define a conserved global current ${\cal J}_{\vec m} \sim *d{\cal A}_{\vec m}$.
Thus, the C-S gauge theory possesses $U(1)^{b_2}$ global symmetry. Using the equation of motion for ${\cal A}_{\vec m}$, one can write ${\cal J}_{\vec m}$ in terms of the bi-fundamental superfields in the quiver gauge theory.

The gauge fields ${\cal A}_{\vec m}$ are reminiscent of the ``statistics gauge fields'' for quasi-particles in the effective description of the fractional quantum Hall effect (FQHE) \cite{Arovas:1985yb} (for a review, see \cite{Wensbook}).  If ${\cal A}$ is one of these $U(1)^{b_2}$ gauge fields for which the Chern-Simons term in the action is
\es{CSTermA}{
{k \over 4 \pi} \int {\cal A} \wedge d {\cal A} \,,
}
the equation of motion for ${\cal A}$ implies that an excitation with charge $q$ under ${\cal A}$ is also a vortex with $2\pi q/k$ units of magnetic flux.  Interchanging two such vortices results in an additional phase
\es{Phase}{
\Delta \phi = \pi {q^2 \over k} \,,
}
showing that the coupling to ${\cal A}$ may change the statistics of the excitations that couple to this gauge field.  This situation is reminiscent of the effective description of the FQHE at filling fraction $1/k$ where quasi-particles have non-trivial statistics due to coupling to a Chern-Simons gauge field.

However, our construction differs in an important way from
standard FQHE systems because we are studying conformal Chern-Simons gauge theories coupled to massless
scalars and fermions. Instead of massive quasi-particles we can only talk about quasi-particle
creation operators (a term recently coined for this situation is ``quasi-unparticles'' \cite{Faulkner:2010tq}). Such operators create vortices that contain the C-S magnetic fluxes and are therefore known as monopole operators. Instead of the diagonal magnetic flux $d{\cal A}_+$, which is known to correspond to the Kaluza-Klein charge in M-theory \cite{Aharony:2008ug}, these operators excite the $b_2$ non-diagonal monopole fields $d{\cal A}_{\vec m}$. Thus, the non-diagonal monopole operators are the only objects that are charged under the $U(1)^{b_2}$ global symmetry of the C-S gauge theory. It has been argued quite convincingly that the M-theory objects dual to such non-diagonal monopole operators are the M2-branes wrapping some of the $b_2$ topologically non-trivial cycles \cite{Imamura:2008ji,Imamura:2009ur}.

The dimensions of the monopole operators in the non-interacting diagonal $U(1)$ have been studied in \cite{Gaiotto:2008ak,Gaiotto:2009tk,Benna:2009xd} following \cite{Borokhov:2002cg}, but the dimensions of the ``non-diagonal'' monopole operators appear to be harder to calculate on the gauge theory side.  The AdS/CFT correspondence predicts that the dimensions of the operators dual to the wrapped M2-branes scale as $\sqrt N$ for large $N$, but presumably, this is difficult to test. Nevertheless, if we simply accept the proposal of \cite{Imamura:2008ji,Imamura:2009ur}, we find an interesting picture where the $b_2$ topological wrapped M2-brane charges in $AdS_4\times Y^7$ are mapped to the $b_2$ $U(1)$ global charges in the dual quiver Chern-Simons gauge theory. In particular, a uniform density of such a topological charge corresponds to a uniform $U(1)$ magnetic field in the C-S gauge theory.  The magnetic field here is not quite the same as in the duals to the dyonic black holes of \cite{Hartnoll:2007ip}, where the magnetic field was added as an external background. We may nevertheless speculate that the zero-temperature entropy of our topologically charged brane solution is due to the degeneracy of Landau levels on the gauge theory side.

\subsection{Boundary conditions in $AdS_4$ and wrapped branes}

In the AdS/CFT correspondence, a conserved current of a field theory is mapped to a massless gauge field in the bulk.
The gauge fields corresponding to the conserved currents ${\cal J}_{\vec m}$ are then the $A^{(i)}$, $i = 1,\ldots, b_2$, that enter the fluctuation $\delta A_3$ in (\ref{Top2Currents}). An additional phenomenon special to $AdS_4$ is that the dual gauge fields $\tilde A^{(i)}$ in (\ref{Top2Currents}) correspond to the C-S gauge fields
${\cal A}_{\vec m}$ in the gauge theory.  Indeed, as shown in \cite{Witten:2003ya}, any two gauge fields $\tilde A$ and $A$ in $AdS_4$ that satisfy $d\tilde A = *_4 dA$ should be quantized so that one corresponds to a gauge field ${\cal A}$ in the dual field theory and the other one to the dual conserved current ${\cal J} = *_3 d{\cal A}$.

Let us write the $AdS_4$ metric in the form
\es{AdS4Metric}{
ds^2 = {1 \over z^2} \left(-dt^2 + d\vec{x}^2 + dz^2 \right) \,,
}
and pass to a gauge where $\tilde A_z = A_z = 0$.  Near $z = 0$, the fields $\tilde A$ and $A$ have the following expansion
\es{BoundaryExpansion}{
\tilde A &= \tilde a^{(0)}_m dx^m + z \tilde a^{(1)}_m dx^m + {\cal O}(z^2 \log z) \,, \\
A &=  a^{(0)}_m dx^m + z a^{(1)}_m dx^m + {\cal O}(z^2 \log z) \,.
}
The duality relation between $A$ and $\tilde A$ implies that
\es{DualityRelations}{
da^{(0)} = *_3 \tilde a^{(1)}\,, \qquad d\tilde a^{(0)} = *_3 a^{(1)} \,.
}
Without loss of generality, let us assume that $A$ is dual to the conserved current ${\cal J}$.  This means that $a^{(0)}$ should be interpreted as an external source for ${\cal J}$, while $a^{(1)}$ as the expectation value of ${\cal J}$ (up to normalization).  Adding an external source $a^{(0)}$ for ${\cal J}$ means that the action changes by
\es{deltaS}{
\delta S &= \int d^3 x\, \sqrt{-g}\, a^{(0)}_m {\cal J}^m = \int a^{(0)} \wedge *_3 {\cal J}
= \int da^{(0)} \wedge {\cal A} \\
&=  \int *_3 \tilde a^{(1)} \wedge {\cal A} = \int d^3 x\, \sqrt{-g}\, \tilde a^{(1)}_m {\cal A}^m \,,
}
where we integrated by parts and used \eqref{DualityRelations}.  Equation \eqref{deltaS} shows that if $a^{(0)}$ is an external source for ${\cal J} = *_3 {\cal A}$, then $\tilde a^{(1)}$, which is related to $a^{(0)}$ through \eqref{DualityRelations}, is an external source for ${\cal A}$. So indeed, if $A$ is dual to ${\cal J}$ then $\tilde A$ is dual to ${\cal A}$, provided $d\tilde A = *_4 dA$ and ${\cal J} = *_3 d{\cal A}$.  Similarly, if we assumed that $A$ was dual to ${\cal A}$ we would conclude that $\tilde A$ should be dual to ${\cal J} = *_3 {\cal A}$.

There are thus two possible boundary conditions for the Abelian gauge fields $A$ and $\tilde A$ in $AdS_4$. From now on we will assume that $A$ is one of the topological gauge fields $A^{(i)}$ appearing in the expression for $\delta A_3$, while $\tilde A$ is its dual, as in \eqref{Top2Currents}.  The first (and the more conventional) choice of boundary conditions corresponds to fixing the boundary value of $A$ but allowing the boundary value of $\tilde A$ to fluctuate. With this choice, the M2-branes wrapping a certain two-cycle are gauge invariant because they couple electrically to the gauge field $A$ that vanishes at the conformal boundary, but the M5-branes wrapping the dual cycle are not.  This statement may seem puzzling, but it agrees with the gauge non-invariance of the baryonic operators in the dual C-S gauge theory \cite{Imamura:2008ji}. Indeed, operators of the form $\det X$ where $X$ is one of the bi-fundamental fields are not invariant under the $U(1)$ subgroups of the $U(N)^{2+b_2}$ gauge group. Another choice of boundary conditions corresponds to fixing the boundary value of $\tilde A$ but allowing the boundary value of $ A$ to fluctuate. Now the wrapped M5-branes are gauge invariant, while the wrapped M2-branes are not.  This choice should correspond not to the $U(N)^{2+b_2}$ Chern-Simons gauge theories, but to their appropriate Legendre transforms \cite{Klebanov:1999tb,Witten:2003ya} that turn the $U(N)$'s into $SU(N)$'s.\footnote{We are grateful to D. Jafferis for discussions on this issue.} In the Legendre transformed theories, baryonic operators like $\det X$ are fully gauge invariant, while it is no longer possible to write down non-diagonal monopole operators that correspond to wrapped M2-branes.

\subsection{An example: $AdS_4\times M^{1,1,1}/\Z_{k}$}

\begin {figure}
\center\includegraphics [width=0.25\textwidth] {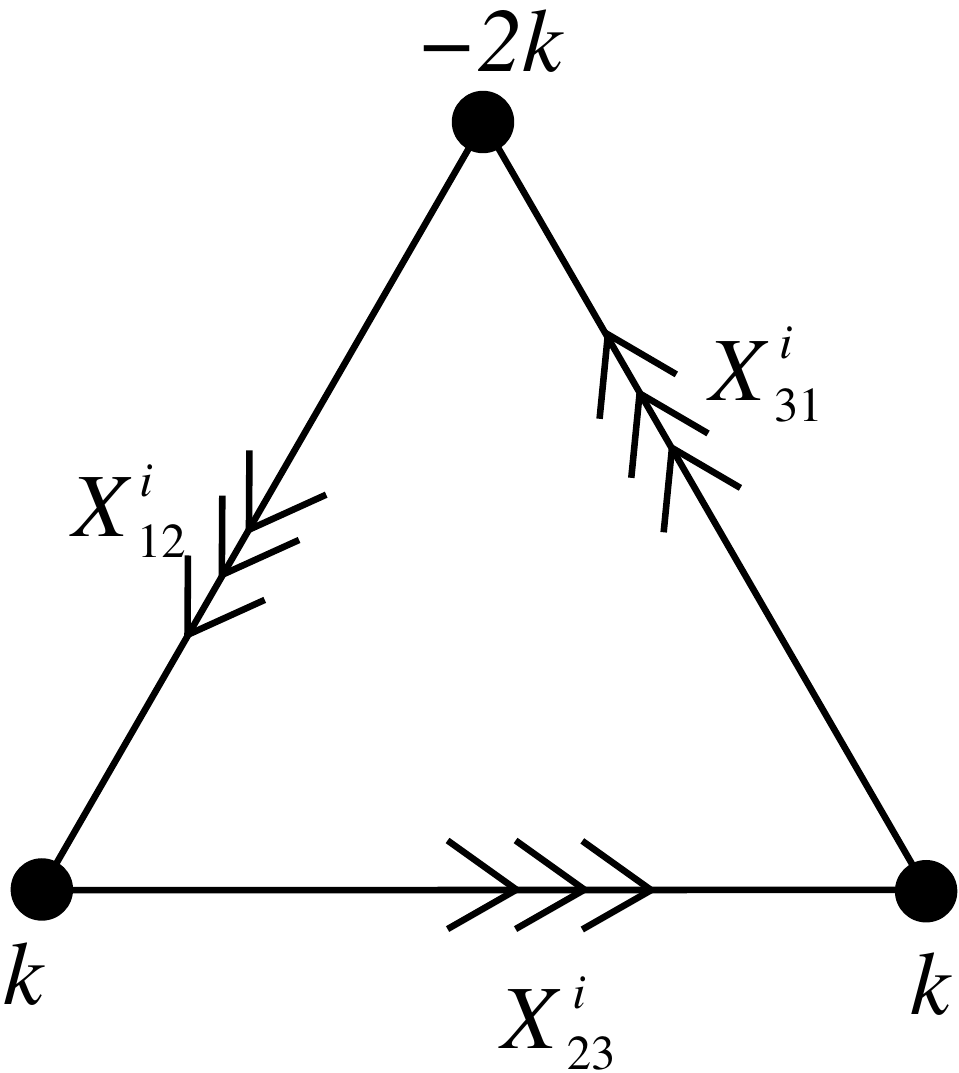}
\caption {The quiver diagram for the C-S gauge theory dual to $AdS_4 \times M^{1, 1,1}/\Z_k$ as conjectured in \cite{Martelli:2008si,Hanany:2008cd}.  The numbers next to the gauge nodes represent the C-S levels.\label {fig:M111quiver}}
\end {figure}

The theory conjectured to be dual to M-theory on $AdS_4\times M^{1,1,1}/\Z_{k}$ \cite{Martelli:2008si,Hanany:2008cd} is the ${\cal N}=2$ superconformal $U(N)_1\times U(N)_2\times U(N)_3$ C-S gauge theory with levels $(-2k,k,k)$ coupled to three sets of bifundamental chiral superfields $X_{12}^i, X_{23}^i, X_{31}^i$, $i=1,2,3$ (see figure~\ref{fig:M111quiver}).
The $SU(3)\times U(1)_R$ invariant superpotential is
\es{superM}{ W\sim \epsilon_{ijk} {\rm Tr} (X_{12}^i X_{23}^j X_{31}^k)
\,.
}
The level assignments break the $\Z_3$ symmetry of the quiver diagram, and the R-charges of the chiral superfields are taken to be \cite{Davey:2009qx}
$R(X_{12})= R(X_{31})=7/9$, $R(X_{23})=4/9$.
The natural way to combine the three $U(1)$ gauge fields is
\es{combinations}{{\cal A}_+ = {\cal A}_1 + {\cal A}_2 + {\cal A}_3\,,
\qquad
{\cal A}_b =  -2 {\cal A}_1 +{\cal A}_2 + {\cal A}_3\,, \qquad
{\cal A}=\sqrt 2 ( {\cal A}_2- {\cal A}_3)\ .
}
The gauge field ${\cal A}$ has the standard Chern-Simons term \eqref{CSTermA}, and it also enters the covariant derivatives for bi-fundamental fields.
Therefore, the ${\cal A}$ equation of motion is
\es{eofmotion}{
{k\over 2\pi} \epsilon^{\mu\nu\lambda} \partial_\nu {\cal A}_\lambda = {\cal J}^\mu \,,
}
where ${\cal J}^\mu$ is the $U(1)$ current
\es{GotJ}{
{\cal J}_\mu &\sim {i \over 2} \tr \left[\bar X_{12}^i D_\mu X_{12}^i
+ \bar X_{31}^i D_\mu X_{31}^i -2 \bar X_{23}^i D_\mu X_{23}^i  \right] + \text{c.c.} + \text{fermionic terms}\,,
}
and $D_\mu$ is the gauge covariant derivative acting on the bi-fundamental fields $X_{ab}^i$ in the fundamental of $U(N)_a$ and anti-fundamental of $U(N)_b$.
The manifold $M^{1, 1, 1}/\Z_k$ has $b_2=1$, and there is one topological $U(1)$ gauge field in $AdS_4$. In the C-S gauge theory, the current dual to it is ${k\over 2\pi} \epsilon^{\mu\nu\lambda} \partial_\nu {\cal A}_\lambda$.

Some results on matching of the chiral operators in this gauge theory with supergravity fluctuations are available \cite{Franco:2009sp}, but none of these operators carry the topological $U(1)$ charge. To construct the operators corresponding to the wrapped M2-branes one has to include the monopole operators with the magnetic flux for the field ${\cal A}$. If we place a unit charge at the origin, $J^0= \delta^2(x)$, then (\ref{eofmotion}) requires that
${\cal A}_\phi={1\over k r}$. This azimuthal gauge field produces phase $2\pi/k$ when another unit charge circles the one
at the origin. This simple field theory argument thus predicts the existence of fractional statistics.
It would be interesting to study how this effect arises for wrapped M2-branes in $AdS_4\times M^{1,1,1}/\Z_{k}$,
but we leave this for future work. We further note that a brane carrying a uniform topological charge density  corresponds in the $U(N)^3$ gauge theory described above to the presence of a constant magnetic field $d{\cal A}$. The ground state of the charged fields in this background is expected to exhibit the Landau level degeneracy. It would be interesting to investigate if this degeneracy may help explain the large $T=0$ entropy found on the gravity side.

As reviewed above, the standard boundary conditions in $AdS_4$ allow the wrapped M2-branes but make the wrapped M5-branes transform under the corresponding $U(1)$ gauge transformations \cite{Imamura:2008ji,Imamura:2009ur}. This agrees with the fact that operators like ${\rm det} X_{23}$ transform under the ${\cal A}$ gauge transformations in the $U(N)^3$ gauge theory. One can, however, change the $AdS_4$ boundary conditions to make the wrapped M5-branes allowed and M2-branes forbidden. The corresponding operation in the gauge theory is a Legendre transform \cite{Klebanov:1999tb,Witten:2003ya}, which turns the $U(1)$ into a global symmetry. Since the gauge field ${\cal A}$ becomes non-dynamical, we can no longer use monopole operators involving this gauge field; this agrees with the fact that the wrapped M2-branes are not allowed. In the Legendre transformed theory we can, however, write down baryonic operators like ${\rm det} X_{23}$ of dimension $4N/9$. This dimension agrees with the volume of one of the five-cycles in $M^{1,1,1}$  \cite{Fabbri:1999hw}. This discussion of baryonic operators is rather sketchy, and a number of issues remain to be elucidated. In particular, it would be interesting to study the Legendre transformed theory in more detail.

\section{A BTZ black hole in type IIB theory}
\label{DUALITIES}

It is interesting to study a reduction of our M-theory membrane solutions to string theory.
Since all fields are independent of the two spatial directions $x^1$ and $x^2$, we may consider the following strategy.
First, we compactify these directions on circles of radii $R_1$ and $R_2$, respectively. Then we reduce to type IIA string theory along the $x^2$ direction and perform T-duality along the $x^1$ direction to obtain a type IIB background with eight compact dimensions consisting of $S^1$ times a warped $Y^7$, and with warp factors depending on the radial coordinate $r$.
What makes these transformations particularly interesting is that our analytic solution (\ref{AttractorSoln}), which seems to have unacceptable large $r$ behavior in M-theory, acquires conventional $AdS_3$ asymptotics in the type IIB theory. Furthermore, the type IIB background, supported by $F_5$ flux only, turns out to be the product of a squashed $Y^7$ space and an extremal BTZ black hole \cite{Banados:1992wn, Banados:1998gg}.

Some of the reasons for this simplicity can be traced back to our original M-brane construction.
We start with a stack of $N$ M2-branes spanning the $(t, x^1, x^2)$ directions placed at the tip of the cone over $Y^7$, and then add a density of M2-branes wrapping two-cycles inside $Y^7$. Upon reduction to IIA, the $N$ M2-branes wrapping $T^2$ turn into $N$ fundamental strings winding around the $x^1$ circle, while the other wrapped M2-branes turn into wrapped D2-branes. Upon T-duality, the winding modes turn into momentum modes which affect the metric only and do not source the NS-NS two-form $B_2$, while the wrapped D2-branes turn into wrapped D3-branes.
The type IIB background therefore describes D3-branes wrapping a two-cycle in $Y^7$ and a circle, with $N$ units of momentum flowing along the circle. This setup is very similar to the original D-brane constructions of supersymmetric black holes with non-vanishing Bekenstein-Hawking entropy \cite{Strominger:1996sh,Callan:1996dv,Klebanov:1996mh}. For example, one such construction involves two stacks of D3-branes
wrapping two-tori embedded inside $T^6$ and intersecting over a circle, while we instead have D3-branes wrapping more complicated cycles inside a squashed $Y^7$.
As a result, our background does not appear to be supersymmetric.

We give the reduction of our general background \eqref{11dMetric}--\eqref{F4M2} from M-theory to type IIA and the T-duality to type IIB in Appendix~\ref{APPENDIXREDUCE}.   In this section we will restrict our attention to a slightly generalized version of the exact solution from section~\ref{ATTRACTOR}, which we will connect through dimensional reduction and T-duality to a 
locally $AdS_3\times \text{squashed }Y^7$ type IIB background.  We do this starting from the type IIB solution in section~\ref{ADS3SOLUTION}.  In section~\ref{MLIFT} we discuss the corresponding M-theory background.

\subsection{The type IIB background}
\label{ADS3SOLUTION}

\newcommand{\LAdS}{L_3}

Let us start with the following ten-dimensional string frame background describing a product of a locally $AdS_3$ space and a squashed $Y^7$:
\es{MoreGeneralMetric}{
ds_{10}^2 &=  \left[ {r^2 \over \LAdS^2} \left(-d t^2 + d x^2 \right) + {\LAdS^2 \over r^2} d r^2  +
\alpha \left(d t + d x \right)^2 \right] \\
{}&+8 \LAdS^2 \left[ds_{V_1}^2 +
{1 \over 2} ds_{V_2}^2 + {4 \over 3} (d\psi + \sigma_1 + \sigma_2)^2 \right] \,, \\
F_5 &=  8 \sqrt{2 \over 3} r
d t \wedge d x \wedge d r \wedge (2 \omega_1 - \omega_2)
- {512  \over 3} \LAdS^4  \omega_1\wedge (\omega_2 - \omega_1)
\wedge (d\psi + \sigma_1 + \sigma_2) \,,
}
where $\LAdS$ is the radius of the asymptotically $AdS_3$ space and $\alpha$ is an arbitrary constant.  
The Lorentz boosts $x-t\rightarrow \lambda^{-1} (x-t),\ x+t\rightarrow \lambda (x+t)$ act as $\alpha\rightarrow \lambda^2\alpha$; therefore, there are only three distinct cases: $\alpha>0$, $\alpha=0$, and $\alpha<0$. The locally $AdS_3$ space with positive $\alpha$ describes an extremal BTZ black hole \cite{Banados:1992wn, Banados:1998gg}, which has the smallest mass for a given angular momentum.

This IIB background describes a state in the $(1+1)$-dimensional CFT on D3-branes wrapped around the $x$-circle as well as a two-cycle in the internal space.  Not much is known about this gauge theory, but using the gauge/string correspondence one can extract the central charge from the Weyl anomaly \cite{Henningson:1998gx}:
\es{CentralCharge}{
c = {3 \over 2} {\LAdS \over G_3} = 12 \pi {\LAdS \over \kappa_3^2} \,,
}
where $\kappa_3$ is the effective gravitational constant in three dimensions.  The $3$-d gravitational constant can be expressed in terms of the gravitational constant of the type IIB theory, $\kappa_{10}$, through
\es{Gotkappa3}{
{1 \over \kappa_3^2} = 2^{10} \sqrt{2 \over 3} \LAdS^7 \Vol(V_1) \Vol(V_2) \Delta \psi {1 \over \kappa_{10}^2} \,,
}
the factor multiplying $1/\kappa_{10}^2$ in this equation being just the volume of the internal space.

To estimate the number of D3-branes we compute the number of $F_5$ flux units through a non-trivial five-cycle in the internal space. One of the simplest such five-cycles spans  $V_1$ and the fiber direction.  The number of units of D3-brane flux through it can be computed from the standard formulae
\es{D3Flux}{
N_{D3} = {1 \over 2 \kappa_{10}^2 \tau_{D3}} \int F_5 \,,
\qquad \tau_{D3} = {2 \pi \over g_s (2 \pi \ell_s)^4} \,,
\qquad {1 \over 2 \kappa_{10}^2} = {2 \pi \over g_s^2 (2 \pi \ell_s)^8} \,,
}
which give
\es{GotD3Flux}{
N_{D3} = {2^9 \over 3} {\LAdS^4 \over \sqrt{\pi} \kappa_{10}} \Vol(V_1) \Delta \psi \,.
}
Comparing this expression with the one for the central charge above, we notice that $c \sim N_{D3}^2$, suggesting an interpretation of the central charge in terms of intersecting D3-branes.

The gravity background above does not correspond to the vacuum state of the gauge theory---the vacuum has $\alpha = 0$.  Nonzero $\alpha$ translates into a nonzero expectation value of the stress-energy tensor.  The AdS/CFT dictionary gives
\es{GotTmn}{
\langle T_{tt} \rangle = \langle T_{tx} \rangle =
\langle T_{xx} \rangle = {\alpha \over  \kappa_3^2 \LAdS} \,,
}
so in the field theory there is conformal matter moving at the speed of light in the negative $x$ direction.  If we compactify the $x$ direction on a circle of radius $R_x$, the entropy of this state can be computed in gravity from the area of the horizon at $r = 0$:
\es{GotEntropy}{
S = {(2 \pi)^2 \alpha^{1 \over 2} R_x \over \kappa_3^2} \,.
}

There is a way of understanding this entropy from field theory considerations, which provides a consistency check on the above formulas.  Since the $x$ direction is a circle of radius $R_x$, the momentum along it needs to be quantized in units of $1/R_x$.  The number of momentum units is
\es{GotN}{
N = R_x \abs{p_x} =  2 \pi R_x^2 \abs{\langle T_{tx} \rangle} =
2 \pi {\alpha R_x^2 \over  \kappa_3^2 \LAdS} \,.
}
Combining this relation with \eqref{CentralCharge} and \eqref{GotEntropy} we verify the Cardy formula
\es{SNcRelation}{
S = 2 \pi \sqrt{N c \over 6} \,,
}
which can be derived by assuming that the entropy comes from the number of ways of partitioning the $N$ units of momentum into smaller momentum quanta.

\subsection{The dual M-theory background}
\label{MLIFT}

T-dualizing \eqref{MoreGeneralMetric} along the compact direction $x$ and lifting to M-theory by introducing a new coordinate $y$, one obtains the metric
\es{11dMetricMoreGeneral}{
ds_{11}^2 &= h^{-{2 \over 3}} \left[- {r^4 \over \LAdS^4 } d t^2 + d x^2 + d y^2 \right]
+ h^{1 \over 3} {\LAdS^2 \over  r^2}  d r^2 \\
{}&+ 8 h^{1 \over 3} \LAdS^2 \left[ds_{V_1}^2
+ {1 \over 2} ds_{V_2}^2 + {4 \over 3} (d\psi + \sigma_1 + \sigma_2)^2 \right] \,,
\qquad h \equiv {r^2 \over \LAdS^2} + \alpha \,.
}
The four-form $F_4$ is
\es{F4MoreGeneral}{
F_4 &= -2 {r \alpha \over h^2 \LAdS^2}
d t \wedge d x \wedge d y \wedge d r
- 8 \sqrt{2 \over 3} r d t \wedge d r
\wedge (2 \omega_1 - \omega_2)  \,.
}
For $\alpha<0$ the metric contains a naked singularity at finite $r$, while the $\alpha=0$ case also appears to be singular. 
We are therefore primarily interested in the $\alpha>0$ where the M-theory metric is equivalent to \eqref{MetricF4Attractor}.

In going from type IIB to type IIA string theory, the circle of radius $R_x$ gets replaced by a circle of radius $\tilde R_x = \ell_s^2 / R_x$, $\ell_s \equiv \sqrt{\alpha'}$ being the string length.  In addition, the string coupling constant $g_s$ of the type IIB theory becomes $\tilde g_s = g_s \ell_s / R_x$ in type IIA\@.  The lift to M-theory introduces the new compact direction $y$ of radius $\tilde R_y = \tilde g_s \ell_s$ and sets the Planck length in eleven dimensions equal to $\ell_p = \tilde g_s^{1 \over 3} \ell_s$.  The $11$-d gravitational constant $\kappa_{11}$ is related to the gravitational constant $\kappa_{10}$ in the IIB theory by $\kappa_{11}^2 = 2 \pi \kappa_{10}^2 \tilde R_x \tilde R_y / R_x$ as follows from the relations $2 \kappa_{11}^2 = (2 \pi \ell_p)^9 / (2 \pi)$, $2 \kappa_{10}^2 = (2 \pi \ell_s)^8 g_s^2 / (2 \pi)$ and the duality transformations described above.

Using the relations between the various constants in M-theory and type IIB mentioned in the previous paragraph, one can easily check that the Bekenstein-Hawking entropy of the $11$-d black hole in \eqref{11dMetricMoreGeneral} with an event horizon at $r = 0$ agrees precisely with the expression \eqref{GotEntropy} that we found in ten dimensions.   One can also check that the number of M2-branes filling the $(t, x, y)$ directions,
\es{SpaceFilling}{
N = {1 \over 2 \kappa_{11}^2 \tau_{M2}} \int F_7 \,, 
}
agrees with the number of units of momentum in the $x$ direction in the $10$-d background that was computed in eq.~\eqref{GotN}.

It is not hard to check that for $\LAdS = 2^{3 \over 16} 3^{-{9 \over 16}}$ and $\alpha =  2^{- {45 \over 8}} 3^{39\over 8}$, the change of coordinates $r \to 2^{-{21 \over 8}} 3^{15 \over 8}\sqrt{r^4 - 1}$, $t \to 2^{{21 \over 4}} 3^{-{19 \over 4}} e^{-{1 \over 2} w_0} t$, $x \to {3 \over 2} x^1$, and $y \to {3 \over 2} x^2$ brings the eleven-dimensional metric \eqref{11dMetricMoreGeneral} into the form of the exact solution \eqref{MetricF4Attractor}.  When the size of the torus parameterized by $x$ and $y$ in eleven dimensions is small in Planck units, one can thus view the effective IR theory described by the attractor solution \eqref{AdS2Solution}---which is the IR limit of \eqref{MetricF4Attractor}---as defined through the asymptotically $AdS_3$ background in type IIB theory that we discussed above. In this limit, one can argue that at nonzero charge density the effective IR description of the $(2+1)$-dimensional C-S gauge theory dual to $AdS_4 \times Y^7$ is the same as that of a chiral sector of a $(1+1)$-dimensional CFT dual to $AdS_3 \times \text{squashed } Y^7$.

\section{Discussion}

We have constructed new charged membrane backgrounds of M-theory that are asymptotic to $AdS_4\times Y^7$ where $Y^7$ is a Sasaki-Einstein manifold with non-vanishing $b_2$. In particular, we considered $Y^7$ that is a circle bundle over a product of two K\"ahler-Einstein manifolds, $V_1\times V_2$. Instead of the $U(1)_R$ charge corresponding to translations of the circle, that was used in previous M-theory constructions \cite{Denef:2009tp,Gauntlett:2009dn,Gauntlett:2009bh}, we turned on a ``topological'' charge corresponding to a component of $\delta A_3$ along the universal harmonic form $\omega=\omega_1-2\omega_2$.  As the Hawking temperature of the black membrane horizon is decreased, a $U(1)_R$-charged solution typically undergoes a phase transition due to condensation of charged fields. We showed that such a phase transition does not occur for our topologically charged solutions. At $T=0$ the near-horizon region becomes $AdS_2\times \R^2\times \text{squashed } Y^7$, which signals emergent quantum criticality. This throat region is by itself a solution of the 11-d supergravity equations.

If we compactify the brane coordinates $x^1$ and $x^2$ on a two-torus, then the resulting black hole has two kinds of charge. One of them is proportional to the number of M2-branes wrapping the $T^2$, the other to the number of M2-branes wrapping the two-cycle inside $Y^7$.  To study whether charge condensation occurs, we calculated the potential as a function of $r$ for the different types of wrapped M2-branes. We found that the M2-branes wrapping the internal cycles experience attractive forces at any temperature; the M2-branes wrapped over $T^2$ experience an attractive force that tends to zero as
$T\rightarrow 0$ for all $r$. Thus, unlike the R-charged brane solutions or the type IIB 3-brane solution with a baryonic charge \cite{Herzog:2009gd}, the new M-theory solution does not suffer from an instability with respect to expulsion of toroidal branes to large $r$ \cite{Hartnoll:2009ns,Yamada:2007gb, Yamada:2008em}.

The fact that there is a moduli space for the M2-branes wrapped over $T^2$ is consistent with the conjecture that gravity is the weakest force, which implies that there should be a charged object {\it not} attracted to an extremal charged black hole horizon \cite{ArkaniHamed:2006dz}. Nevertheless, it is very surprising to find the vanishing of the potential for a probe space-time filling M2-brane in a background which apparently does not preserve any supersymmetry. It would be interesting to investigate if this moduli space is lifted by higher-derivative corrections to the 11-d supergravity action, which are expected to correspond to $1/N$ corrections in the dual Chern-Simons gauge theory.

When a string or M-theory background does not preserve any supersymmetry, one should be concerned about various potential instabilities. We have shown that there is no low-temperature condensation of charged objects, but one should also check the perturbative stability of neutral fluctuations. We have carried out some preliminary checks for neutral scalars, but clearly more should be done. Finally, there may be some non-perturbative gravitational instabilities but their study is beyond the scope of this paper.

If our zero-temperature solution is completely stable, we should try to explain the microscopic origin of its large Bekenstein-Hawking entropy. One approach may be to study a dual quiver Chern-Simons gauge theory with a constant background magnetic field $d{\cal A}$ which produces a uniform $U(1)$ global charge density. Our membrane solution implies that this gauge theory develops IR quantum criticality corresponding to the appearance of the $AdS_2$ throat. We would like to gain some understanding of this phenomenon. It would also be interesting to study the apparent fractional statistics of the wrapped M2-branes in the $AdS_4\times M^{1,1,1}/\Z_k$ background.

Another possible microscopic approach to the IR theory is motivated by the type IIB background \eqref{MoreGeneralMetric} which is related by string dualities to the M-theory exact solution \eqref{MetricF4Attractor} with different large $r$ asymptotics. 
In this type IIB background, the near-horizon $AdS_2$ region arises from a reduction of the extremal BTZ black hole on a circle.
The extremal BTZ times squashed $Y^7$ background should be dual to the $(1+1)$-dimensional CFT on D3-branes partially wrapped over the squashed $Y^7$. Calculating the central charge of this CFT would provide a way of explaining the charged black hole entropy via \eqref{SNcRelation}.

\section*{Acknowledgments}
We thank M.~Kiermeier for collaboration in the early stages of this project, and J.~Gauntlett, C.~Herzog, D.~Jafferis, J.~Polchinski, M.~Rangamani, E.~Silverstein, P.~Yang, and A.~Yarom for useful discussions.  SSP is grateful to A.~Caraiani and S.~Sivek for answering patiently so many questions on complex algebraic geometry. This work was supported in part by the US NSF  under Grant No. PHY-0756966.


\appendix

\section {Metrics for the regular Sasaki-Einstein spaces}
\label {METRICS}

In this section we give the explicit metrics for the regular Sasaki-Einstein manifolds described in section~\ref {EXAMPLES}. We also discuss the non-trivial cycles found in the bases of these manifolds that are useful for the probe brane computations in section~\ref{M2GENERALWRAPPED} and Appendix~\ref{APPENDIXWRAPPED}.

Let us first describe the general approach to computing the range of the coordinate $\psi$ appearing in \eqref{YMetric}.  When the scalars $\chi$ and $\eta_i$ vanish identically, \eqref{11dMetric} solves the eleven-dimensional SUGRA equations with $M = AdS_4$.  Insisting that the radius of $AdS_4$ should be $L$, one finds that the metric on $V_i$, $i = 1, 2$, should be normalized so that the curvature two-form is related to the K\"ahler form through $R_i = 8 \omega_i$.  By definition, the first Chern class of $V_i$ is $c_1 (V_i) \equiv {1 \over 2 \pi} R_i$, so
\es{GotFirstChern}{
c_1(V_i) = {4 \over \pi} \omega_i = {2 \over \pi} d \sigma_i \,,
}
where in the second equality we used \eqref{sigmaKahler}.  Note that $c_1(V_i)$ does not depend on the overall normalization of the metric on $V_i$, but of course the proportionality constant between $c_1(V_i)$ and $\omega_i$ does.  By definition, the first Chern class of the fiber bundle $Y^7 \to V_1 \times V_2$ is
\es{FirstChernFibration}{
c_1 = {1 \over L_f} d(\sigma_1 + \sigma_2) \,,
}
where $L_f$ is the length of the fiber, i.e.~the range of $\psi$.  From comparing \eqref{FirstChernFibration} to \eqref{GotFirstChern} we see that in order to compute the length of the fiber we need to know the relation between the first Chern class $c_1$ of the fiber bundle and the first Chern class $c_1 (V_1) + c_1 (V_2)$ of the base.  Using a Thom-Gysin sequence, one can show \cite{Friedrich:1990zg, Martelli:2006yb} that the only requirement is that $c_1(V_1) + c_1 (V_2)$ should be an integer multiple of $c_1$.  Recalling that $c_1$ and $c_1(V_i)$ represent cohomology classes with integer coefficients, we denote by $a_i$ the largest integers so that ${1 \over a_i} c_1 (V_i) \in H^2(V_i; \Z)$.  Since $c_1(V_1) + c_1 (V_2)$ must be an integer multiple of $c_1$, one can take
\es{Twoc1s}{
c_1 = {1 \over a} \left(c_1 (V_1) + c_1(V_2) \right) \,,
}
where $a$ can be any common divisor of $a_1$ and $a_2$.  The length of the fiber is then
\es{GotLf}{
L_f = {\pi \over 2} a \,.
}

Seven-dimensional Sasaki-Einstein spaces like the ones above have ${\cal N} = 2$ supersymmetry.  The two Killing spinors are proportional to $e^{\pm 2 i \psi}$ \cite{Martelli:2006yb}, and they are well-defined as long as the range of $\psi$ is an integer multiple of $\pi/2$.  Equation \eqref{GotLf} shows this is indeed the case.

\subsection{$M^{1, 1, 1}$}
\label{M111METRIC}

The manifold $M^{1, 1, 1}$ is the homogeneous space ${SU(3) \times SU(2) \times U(1) \over SU(2) \times U(1) \times U(1)}$ and by construction its isometry group is that of the standard model, $SU(3) \times SU(2) \times U(1)$ \cite{Fabbri:1999hw}.  The cone over $M^{1, 1, 1}$ is a Calabi-Yau four-fold that can be described as a K\"ahler quotient $\C^5 // \C^*$ as follows.  One starts with $\C^5$ parameterized by the complex coordinates $(u^1, u^2, u^3, v^1, v^2)$ and endowed with the K\"ahler potential
\es{KahlerPotM111}{
K = \left(2 u^i \bar u_i \right)^{3 \over 4}
\left(3 v^j \bar v_j \right)^{1 \over 2} \,.
}
One then takes the K\"ahler quotient of this space with charges $(2, 2, 2, -3, -3)$, meaning that we restrict our attention to a submanifold of $\C^5$ defined by
\es{M111Cone}{
2 \left(\abs{u^1}^2 + \abs{u^2}^2 + \abs{u^3}^2 \right)
= 3 \left(\abs{v^1}^2 + \abs{v^2}^2\right) \,,
}
which we further mod out by the equivalence relation
\es{EquivRelation}{
u^i \sim e^{2 i \delta} u^i \,,
\qquad v^j \sim e^{-3 i \delta} v^j \,.
}
The space described by equations \eqref{M111Cone} and \eqref{EquivRelation} is precisely the cone over $M^{1, 1, 1}$.  This space is a cone because both of these equations are invariant under $u^i \to \lambda u^i$ and $v^j \to \lambda v^j$ with $\lambda \in \R_+$. One can check that the induced metric coming from the K\"ahler potential \eqref{KahlerPotM111} is Ricci flat, so the cone over $M^{1, 1, 1}$ is indeed Calabi-Yau.  One can check that the holomorphic four-form $\Omega_4$ on the cone is given by
\es{HolFourFormM111}{
\Omega_4 \sim d \Omega_3 \,, \qquad
\Omega_3 \equiv \left( \epsilon_{i_1 i_2 i_3}u^{i_1}  du^{i_2} \wedge du^{i_3} \right)
\wedge \left( \epsilon_{j_1 j_2} v^{j_1} dv^{j_2} \right) \,.
}

The space $M^{1, 1, 1}$ can be obtained by fixing the overall magnitude of $u^i$ and $v^j$:
\es{FixMagnitude}{
2 \left(\abs{u^1}^2 + \abs{u^2}^2 + \abs{u^3}^2 \right)
= 3 \left(\abs{v^1}^2 + \abs{v^2}^2\right) = 1 \,.
}
An explicit Sasaki-Einstein metric can be found from \eqref{KahlerPotM111} by using the parameterization
\begin{align}
u_1 &= {1 \over \sqrt{2}} \sin \mu \cos {\theta_1 \over 2} e^{{i \over 2} (\phi_1 + \psi_1 + R_u \psi)} \,,
&  v_1 &=  {1 \over \sqrt{3}} \cos {\theta_2 \over 2} e^{{i \over 2} (\phi_2 + R_v \psi)}  \,, \notag\\
u_2 &=  {1 \over \sqrt{2}} \sin \mu \sin {\theta_1 \over 2} e^{{i \over 2} (-\phi_1 + \psi_1 + R_u \psi)} \,,
& v_2 &=  {1 \over \sqrt{3}} \sin {\theta_2 \over 2} e^{{i \over 2} (-\phi_2 + R_v \psi)} \label{uvDefs} \,, \\
u_3 &=  {1 \over \sqrt{2}} \cos \mu \, e^{{i \over 2} R_u \psi} \,. \notag
\end{align}
for any $R_u$ and $R_v$ satisfying $3 R_u + 2 R_v = 1$.  This metric has the form \eqref{YMetric} with\footnote{The metric obtained from \eqref{KahlerPotM111} does not depend on the angle $\delta$ appearing in \eqref{EquivRelation}.  One way to see this is to promote \eqref{EquivRelation} to  $u^i \to \lambda^2 u^i$, $v^j \to \lambda^{-3} v^j$  with $\lambda \in \C^*$ and think of \eqref{FixMagnitude} as a gauge fixing condition for this transformation.  Since the K\"ahler potential is independent of $\lambda$, which can be regarded as a complex coordinate in $\C^5$, the metric on $\C^5$ following from \eqref{KahlerPotM111} is degenerate in the $\lambda$ direction.}
\es{V12MetricM111Again}{
ds_{V_1}^2 &= {3 \over 4} \biggl[ d\mu^2 + {1 \over 4} \sin^2 \mu
\left(s_1^2 + s_2^2 + \cos^2 \mu\, s_3^2  \right) \biggr] \,, \\
ds_{V_2}^2 &= {1 \over 8} \biggl[ d \theta_2^2 + \sin^2 \theta_2 d \phi_2^2 \biggr] \,,
}
and
\es{GotSigmaM111Again}{
\sigma_1 = {3 \over 8} \sin^2 \mu\, s_3 \,, \qquad
\sigma_2 = {1 \over 4} \cos \theta_2 d \phi_2 \,.
}
In the above equations we have defined
\es{sDefsAgain}{
s_1 \equiv d \theta_1\,, \qquad
s_2 \equiv \sin \theta_1 d \phi_1\,, \qquad
s_3 \equiv d \psi_1 + \cos \theta_1 d \phi_1 \,.
}
The metrics \eqref{V12MetricM111Again} describe $V_1 = \CP^2$ and $V_2 = \CP^1$.

Let the hyperplane divisor $H$ be the generator of $H_2 (\CP^2; \Z) \cong \Z$.  $H$ is the homology class of a $\CP^1 \subset \CP^2$, so in homogeneous coordinates $H$ can represented by the two-cycle $\left\{ [0,  z_1,  z_2] : z_1, z_2 \in \C \right\}$.  Let us denote by $D$ the generator of $H_2(\CP^1; \Z) \cong \Z$.  From \eqref{GotFirstChern}, one can compute
\es{FirstChernHomM111}{
\int_H c_1(V_1) = 3\,, \qquad \int_D c_1(V_2) = 2 \,,
}
so in this case $a_1 = 3$ and $a_2 = 2$.  There is only one possibility for $a | \gcd(a_1, a_2)$, namely $a = 1$.  The length of the fiber is $\pi / 2$.

\subsection{$Q^{1, 1, 1}$ and $Q^{2, 2, 2}$}
\label{Q111METRIC}

The space $Q^{1, 1, 1}$ is also a homogeneous space, ${SU(2) \times SU(2) \times SU(2) \over U(1) \times U(1)}$ \cite{Fabbri:1999hw}.  The cone over it is Calabi-Yau and can be constructed from taking a K\"ahler quotient $\C^6 // \C^{*2}$.  If the coordinates on $\C^6$ are $(a^1, a^2, b^1, b^2, c^1, c^2)$, the K\"ahler quotient can be thought of as the level sets
\es{LevelSetsQ111}{
\abs{a^1}^2 + \abs{a^2}^2 =
\abs{b^1}^2 + \abs{b^2}^2 =
\abs{c^1}^2 + \abs{c^2}^2 \,,
}
and the following identifications
\es{EquivRelQ111}{
a^i &\sim e^{i \delta} a^i\,,  \qquad b^j \sim e^{-i \delta} b^j \,,  \qquad c^k \sim c^k \,, \\
a^i &\sim e^{i \delta} a^i\,,  \qquad b^j \sim b^j  \,,  \qquad c^k \sim e^{-i\delta} c^k \,.
}
With the K\"ahler potential
\es{KahlerPotQ111}{
K = \left(a^i \bar a_i \right)^{1 \over 2}
\left(b^j \bar b_j \right)^{1 \over 2}
\left(c^k \bar c_k \right)^{1 \over 2} \,,
}
the cone over $Q^{1, 1, 1}$ is Calabi-Yau.  The holomorphic four-form $\Omega_4$ is in this case
\es{HolFourFormQ111}{
\Omega_4 \sim d \Omega_3\,, \qquad
\Omega_3 \equiv \left( \epsilon_{i_1 i_2} a^{i_1} da^{i_2} \right) \wedge
\left( \epsilon_{j_1 j_2} b^{j_1} db^{j_2} \right) \wedge
\left( \epsilon_{k_1 k_2} c^{k_1} dc^{k_2} \right) \,.
}

In order to find a metric on $Q^{1, 1, 1}$ itself, one needs to restrict to the base of the cone by fixing the overall magnitude of $a^i$, $b^j$, and $c^k$:
\es{FixMagQ111}{
\abs{a^1}^2 + \abs{a^2}^2 =
\abs{b^1}^2 + \abs{b^2}^2 =
\abs{c^1}^2 + \abs{c^2}^2 = 1 \,.
}
From \eqref{KahlerPotQ111} one can obtain an explicit metric on $Q^{1, 1, 1}$ using the parameterization
\begin{align}
a^1 &= \cos {\theta_1 \over 2} e^{{i\over 2} (\phi_1 + R_a \psi)} \,,
& a^2 &= \sin {\theta_1 \over 2} e^{{i \over 2} (-\phi_1 + R_a \psi)}\,, \notag \\
b^1 &= \cos {\theta_2 \over 2} e^{{i\over 2} (\phi_2 + R_b \psi)} \,,
& b^2 &= \sin {\theta_2 \over 2} e^{{i \over 2} (-\phi_2 + R_b \psi)} \,, \label{abcDefs} \\
c^1 &= \cos {\theta_3 \over 2} e^{{i\over 2} (\phi_3 + R_c \psi)} \,,
& c^2 &= \sin {\theta_3 \over 2} e^{{i \over 2} (-\phi_3 + R_c \psi)} \,, \notag
\end{align}
for any $R_a$, $R_b$, and $R_c$ such that $R_a + R_b + R_c = 1$.  The metric takes the form \eqref{YMetric} with
\es{V12MetricQ111Again}{
ds_{V_1}^2 &= {1 \over 8} \sum_{i = 1}^2 \biggl[d \theta_i^2 + \sin^2 \theta_i d \phi_i^2  \biggr]\,, \qquad
ds_{V_2}^2 = {1 \over 8} \biggl[d \theta_3^2 + \sin^2 \theta_3 d \phi_3^2 \biggr] \,,
}
and
\es{GotSigmaQ111Again}{
\sigma_1 = {1 \over 4} \left( \cos \theta_1 d\phi_1 + \cos \theta_2 d \phi_2 \right)\,, \qquad
\sigma_2 = {1 \over 4} \cos \theta_3 d\phi_3 \,.
}
The spaces $V_1$ are $V_2$ are in this case $\CP^1 \times \CP^1$ and $\CP^1$, respectively.

Let us denote the two generators of $H_2 (\CP^1 \times \CP^1; \Z) \cong \Z^2$ by $C_1$ and $C_2$ where $C_1$ is the homology class of the first $\CP^1$ factor and $C_2$ of the second. As in the case of $M^{1, 1, 1}$, let us denote the generator of $H_2 (\CP^1; \Z) \cong \Z$ by $D$.   Starting from \eqref{GotFirstChern} it is easy to see that
\es{FirstChernHomQ111}{
\int_{C_1} c_1(V_1) = \int_{C_2} c_1(V_1) = \int_D c_1 (V_2) = 2 \,,
}
so in this case $a_1 = a_2 = 2$.  Therefore there are two possibilities for the integer $a \vert \gcd(a_1, a_2)$: taking $a = 1$ we obtain the space $Q^{1, 1, 1}$, and taking $a = 2$ we obtain $Q^{2, 2, 2}$.  From \eqref{GotLf} we see that the circle fibers of $Q^{1, 1, 1}$ have length $\pi$, while those of $Q^{2, 2, 2}$ have length $\pi / 2$.

\subsection{Circle bundles over $dP_n \times \CP^1$}

The last class of seven-dimensional regular Sasaki-Einstein manifolds comes from principal $U(1)$ fiber bundles over $dP_n \times \CP^1$, $dP_n$ being the $n$th del Pezzo surface, and $3 \leq n \leq 8$.  Topologically, $dP_n$ can be constructed from $\CP^2$ blown up at $n$ generic points.  (The points being generic means that no three points should be collinear and no six points should lie on a conic.)  The del Pezzo's are known to admit K\"ahler-Einstein metrics with positive Ricci curvature
\cite{Tian:1987if, TianAnother}, but unfortunately these metrics are not known analytically.\footnote{See \cite{Doran:2007zn}
where a K\"ahler-Einstein metric on $dP_3$ was computed numerically.}  Despite this fact, we can still describe some of the properties of the corresponding Sasaki-Einstein spaces.

We take $V_1 = dP_n$ and $V_2 = \CP^1$.  The metric on $V_1$ is not known, but the metric on $V_2$ is given by
\es{V2MetricdelPezzo}{
ds_{V_2}^2 =  {1 \over 8} \biggl[ d \theta_2^2 + \sin^2 \theta_2 d \phi_2^2 \biggr] \,,
}
and
\es{sigmaTwoDelPezzo}{
\sigma_2 = {1 \over 4} \cos \theta_2 d \phi_2 \,,
}
as in the previous two cases.

The second homology group of $dP_n$ is $H_2(dP_n; \Z) \cong \Z^{n+1}$, so there are $n+1$ generators which we will denote by $H$ and $E_i$, $1 \leq i \leq n$.   In algebraic geometry language, $H$ is a hyperplane divisor and $E_i$ are the exceptional divisors of the blown-up points.  As in the previous two sections, we denote by $D$ the generator of $H_2(\CP^1; \Z)$.  Using algebraic geometry, one can show
\es{FirstChernHomdP}{
\int_H c_1(V_1) = 3\,, \qquad \int_{E_i} c_1(V_1) = 1\,, \qquad
\int_D c_1(V_2) = 2 \,.
}
It follows that $a_1 = 1$ and $a_2 = 2$, so again the only possible value of $a$ is $a = 1$, giving fibers of length $\pi/2$.

\section{Other supergravity fluctuations around the exact solution}
\label{FLUCTAPPENDIX}

It might be interesting to consider supergravity fluctuations around the extremal solution found in section~\ref{ATTRACTOR} and see whether these fluctuations cause a run-away instability.  Let us focus on fluctuations depending only on the radial variable $r$.  They typically satisfy second order differential equations whose solutions near the extremal horizon behave as $(r-1)^\alpha$.  The exponent $\alpha$ can be either real or complex.  When it is real, the corresponding fluctuations correspond to either a source or a VEV of an operator in the effective quantum mechanics.  When it is complex, the corresponding fluctuations are oscillatory as a function of $r$ and typically cause an instability.

Investigating the behavior of supergravity fluctuations near the extremal horizon is a hard task, because these fluctuations depend on the details of the Sasaki-Einstein spaces $Y^7$.  We will only examine a particularly simple fluctuation in the case $Y^7 = Q^{1, 1, 1}$.  For $Q^{1,1, 1}$, $V_1 = \CP^1 \times \CP^1$, and the background \eqref{11dMetric}--\eqref{GotF7} is symmetric under interchanging the two $\CP^1$ factors.  The mode that we will look at is the leading $\Z_2$-odd mode that changes the sizes of the two $\CP^1$'s.  We call this mode $\lambda$.

There is a non-linear consistent truncation that includes this additional mode $\lambda$.  The eleven-dimensional metric is
\es{Q111MetricPert}{
ds^2 &= e^{-7\chi/2} ds_M^2 + {1 \over 2} L^2 e^\chi \biggl[
e^{\eta_1 + \lambda} (d \theta_1^2 + \sin^2 \theta_1 d\phi_1^2)
+ e^{\eta_1 - \lambda} (d \theta_2^2 + \sin^2 \theta_2 d\phi_2^2) \\
{}&+ e^{\eta_2} (d \theta_3^2 + \sin^2 \theta_3 d\phi_3^2)
\biggr]
+ {1 \over 4} L^2 e^{\chi - 4 \eta_1 - 2 \eta_2}
(d\psi + \cos \theta_1 d\phi_1 + \cos \theta_2 d\phi_2 +
\cos \theta_3 d\phi_3)^2
}
and the four-form is
\es{Q111F4Pert}{
F_4 &= -{3\over L} e^{-{21\over 2} \chi} \vol_M
+ Q L^3 {e^{-{w \over 2}  - {3 \over 2} \chi} \over r^2} dt \wedge dr \wedge
\biggl[ e^{ 2 \eta_1 + 2 \lambda} \sin \theta_1 d \theta_1 \wedge d\phi_1 \\
{}&+ e^{2 \eta_1 - 2 \lambda} \sin \theta_2 d \theta_2 \wedge d\phi_2
- 2 e^{2 \eta_2} \sin \theta_3 d \theta_3 \wedge d\phi_3
\biggr]   \,.
}
When $\lambda = 0$, equations \eqref{Q111MetricPert} and \eqref{Q111F4Pert} reduce to equations \eqref{11dMetric} and \eqref{F4M2}, respectively.

The linearized equation for $\lambda$ following from the eleven-dimensional supergravity equations of motion is (we set $L = 1$)
\es{lambdaLin}{
\lambda'' + \lambda' \left({2 \over r} + {g' \over g} - {w' \over 2} \right)
+ \lambda {2 e^{-6 \eta_1 - 2 \eta_2 - {9 \over 2} \chi} \over r^4 g}
\left[2 e^{5 \eta_1 + 2 \eta_2} r^4 - r^4  -4 e^{8 \eta_1 + 2 \eta_2 + 3 \chi} Q^2 \right] = 0\,.
}

When evaluated on the extremal solution \eqref{AttractorSoln}, equation \eqref{lambdaLin} has analytical solutions:
\es{lambdaAnalytical}{
\lambda = c_+ (r^4 - 1)^{-{1\over 2} \left(1 + \sqrt{17 \over 3} \right)}
+ c_- (r^4 - 1)^{-{1\over 2} \left(1 - \sqrt{17 \over 3} \right)} \,.
}
In the effective quantum mechanics, the solutions multiplying $c_+$ and $c_-$ correspond respectively to a source and a VEV of an operator of dimension $\Delta = {1 \over 2} \left(1 + \sqrt{17 \over 3} \right) \approx 1.69$.  Since the exponents of $r^4 - 1$ are real, we conclude that these fluctuations do not cause an instability.

\section {Comments on wrapped branes}
\label{APPENDIXWRAPPED}

In this section we tie up some loose ends from our discussion in section~\ref{M2GENERALWRAPPED} of M2-branes wrapping an internal two-cycle in $\tilde Y^7$.  We first discuss in section~\ref{APPENDIXCYCLES} some topological properties of two-cycles in a general Sasaki-Einstein manifold $Y^7$ whose K\"ahler-Einstein base is $V_1 \times V_2$.  In section~\ref{APPENDIXCALIB} we give a proof of the bound \eqref{VolumeIneqSimp} on the volumes of the two-cycles of $\tilde Y^7$.

\subsection{The second homology of $Y^7$}
\label {APPENDIXCYCLES}

Topologically, two-cycles in $Y^7$ are classified by the second homology of $Y^7$ with integer coefficients, $H_2(Y^7; \Z)$.  The homology of $Y^7$ can be calculated from the homology of the base of the fibration, the product manifold $V_1 \times V_2$.  In turn, the homology of $V_1 \times V_2$ can be computed from the homology of $V_1$ and that of $V_2$.   For all of the regular Sasaki-Einstein spaces we are interested in, $V_2 = \CP^1$ and the generator of $H_2(V_2; \Z)\cong \Z$ is represented by $V_2$ itself.  Let us call this generator $D$.  The homology of the K\"ahler-Einstein spaces $V_1$ is in all cases of interest $H_2(V_1; \Z) \cong \Z^k$ and let us denote its generators by $C_i$ with $1 \leq i \leq k$.  We have $k = 2$ for $V_1 = \CP^1 \times \CP^1$; $k=1$ for $V_1 = \CP^2$; and $k = n+1$ for $V_1 = dP_n$.  We pick the orientations of $C_i$ and $D$ so that they can be represented by holomorphic surfaces as opposed to antiholomorphic ones.  The second homology of $V_1 \times V_2$ is then $H_2 (V_1 \times V_2 ; \Z) \cong \Z^{k+1}$, and its generators are constructed as follows.  Given a surface that represents $C_i$ in $V_1$ we can take the direct product between this surface and a point in $V_2$; this product is a closed surface in $V_1 \times V_2$ and represents a generator of $H_2(V_1 \times V_2; \Z)$.  Similarly, the direct product between $V_2$ and a point in $V_1$ is also a closed surface in $V_1 \times V_2$ representing a generator of $H_2(V_1 \times V_2 ; \Z)$.  By abuse of notation we will denote the first $k$ generators of $H_2(V_1 \times V_2 ; \Z)$ by $C_i$ and the $(k+1)$th one by $D$, as they are constructed from the corresponding generators of $H_2(V_1; \Z)$ and $H_2(V_2; \Z)$ in a straightforward way.

It turns out that if $H_2(V; \Z) \cong \Z^{k+1}$ then $H_2(Y^7; \Z) \cong \Z^k$.  The reason why $H_2(Y^7; \Z)$ is smaller than $H_2(V; \Z)$ is that whereas all topologically non-trivial closed surfaces in $Y^7$ project down to topologically non-trivial closed surfaces in $V$, not every closed surface in $V$ can be lifted to a closed surface in $Y^7$. In fact, any two-dimensional surface $S$ in $V$ can be lifted to a {\em three}-dimensional surface $\tilde S$ in $Y^7$ by restricting the circle fibration over $V$ to a circle fibration over $S$. In order for a two-dimensional closed surface $S$ in $V$ to be liftable to a {\em two}-dimensional closed surface in $Y^7$, one has to specify what the fiber coordinate $\psi$ should be at each point in $S$.  There is a topological restriction on the types of closed surfaces $S$ one can lift precisely because it may be impossible to specify consistently what $\psi$ is at all points of $S$.  In algebraic topology language, a consistent assignment of $\psi$ to every point in $S$ gives a global section of the pull-back bundle $\tilde S$, and it is known that any circle bundle, in particular $\tilde S$, admits a global section if and only if it is trivial.  Since circle bundles are completely classified by their first Chern class (the cohomology class of the curvature of the $U(1)$ fibration), it follows that a closed surface $S$ in $V$ is liftable to $Y^7$ if and only if the first Chern class of the circle bundle $\tilde S$ (which is nothing but the pull-back of the first Chern class of $Y^7$ to $S$) is zero in cohomology.  In particular,
\es{LiftableCondition}{
\text{$S$ is liftable} \Longleftrightarrow
\int_S c_1 = 0 \,,
}
where $c_1$ is the first Chern class of $Y^7$.  The above argument works only in the case where the surface $S$ is connected---if $S$ is not connected, then the condition \eqref{LiftableCondition} should be satisfied for each connected component separately.

Equation \eqref{LiftableCondition} suggests\footnote{The following argument is not intended to be a proof.  One can prove the result \eqref{HomCondition} using a Gysin sequence.  See \cite{Fabbri:1999hw} for the cases $Y^7 = Q^{1, 1, 1}$ and $Y^7 = M^{1, 1, 1}$.} how to construct $H_2(Y^7; \Z)$ given $H_2(V; \Z)$: $H_2(Y^7; \Z)$ is isomorphic to the kernel of the map that assigns to each element $C$ in $H_2(V; \Z)$ the integer $\int_C c_1$.  In other words, if we parameterize the homology classes in $H_2(V; \Z)$ by
\es{HomV}{
C = \sum_{i = 1}^k \alpha_i C_i + \beta D \,,
}
with $\alpha_i, \beta \in \Z$, then there is a one-to-one correspondence between elements of the homology $H_2(Y^7; \Z)$ of the total space $Y^7$ and classes $C$ in the homology $H_2(V; \Z)$ of the base $V$ satisfying
\es{HomCondition}{
\sum_{i = 1}^k \alpha_i \int_{C_i} c_1 + \beta \int_D c_1 = 0 \,.
}
Such classes form a $\Z^{k}$ subspace of $H_2(V; \Z) \cong \Z^{k + 1}$, so indeed $H_2(Y^7; \Z) \cong \Z^k$.  Note that only connected surfaces representing $C$ can be lifted to $Y^7$ as embedded closed surfaces, as discussed above.

The first Chern class of the fibration, $c_1$, is by definition the cohomology class of the curvature of the connection one-form $\sigma_1 + \sigma_2$ appearing in the metric \eqref{YMetric}.  By equation \eqref{sigmaKahler}, $c_1$ is proportional to the sum $\omega_1 + \omega_2$ of the K\"ahler forms on $V_1$ and $V_2$, so equation \eqref{HomCondition} becomes
\es{HomConditionAgain}{
\sum_{i = 1}^k \alpha_i \int_{C_i} \omega_1 + \beta \int_D \omega_2 = 0 \,.
}
What's nice about this equation is that since $V_1$ and $V_2$ are Einstein spaces, the integrals of the K\"ahler forms over the cycles $C_i$ and $D$ are topological invariants that are known even when an explicit Einstein metric on $V_1$ or $V_2$ is not known.

As an example, for $Y^7 = M^{1, 1, 1}$, $V_1 = \CP^2$, $V_2 = \CP^1$, and the dimension of $H_2(V_1; \Z)$ is $k = 1$.  Algebraic geometry arguments combined with the condition for an Einstein metric (see also Appendix~\ref{M111METRIC}) give $\int_{C_1} \omega_1 = {3 \pi \over 4}$ and $\int_D \omega_2 = {\pi \over 2}$.  Equation \eqref{HomConditionAgain} shows that the generator of the homology of $Y^7$ has $\alpha_1 = 2$ and $\beta = -3$.  An explicit cycle representing this homology class is given in \eqref{CM111}.

As another example, for $Y^7 = Q^{1, 1, 1}$, $V_1 = \CP^1 \times \CP^1$, $V_2 = \CP^1$, and $k = 2$.  In this case, $\int_{C_1} \omega_1 = \int_{C_2} \omega_1 = \int_D \omega_2 = {\pi \over 2}$.  The second homology of $Y^7$ is therefore generated by $(\alpha_1, \alpha_2, \beta) = (1, -1, 0)$ and $(\alpha_1, \alpha_2, \beta) = (1, 0, -1)$.  Explicit cycles representing these homology classes are given in \eqref{C1Q111}--\eqref{C2Q111}.

As a last comment, note that the above discussion does not change if we replace $Y^7$ by $\tilde Y^7$ because the curvature of the $U(1)$ fibration stays unchanged.  Moreover, any cycle ${\cal C}$ in $\tilde Y^7$ should satisfy~\eqref {LiftableConditionSimp} because ${\cal C}$ is in the same homology class as a two-cycle ${\cal C}'$ constructed by lifting a closed surface $S$ in $V$, and for ${\cal C}'$ equation \eqref{LiftableConditionSimp} is equivalent to \eqref{LiftableCondition}.

\subsection{A lower bound on the volumes of closed two-surfaces in $\tilde Y^7$}
\label {APPENDIXCALIB}

The bound \eqref{VolumeIneqSimp} can be proven by finding a calibration.  A calibration (for two-dimensional surfaces) is a closed two-form $\Omega$ with the property that for any orthonormal tangent vectors $u$ and $v$
\es{CalibrationDef}{
\Omega(u, v) \leq 1 \,.
}
Consequently, the volume of any closed two-dimensional surface ${\cal C}$ in $Y^7$ satisfies
\es{CalibrationVolume}{
\Vol({\cal C}) \geq \int_{\cal C} \Omega \,.
}
Since $\Omega$ is closed, the right-hand side of \eqref{CalibrationVolume} depends only on the homology class of ${\cal C}$.  In the space $\tilde Y^7$ with the metric \eqref{tildeYMetric} we will show that
\es{GotCalibration}{
\Omega = s e^{\chi + \eta_1} \omega_1 + t e^{\chi + \eta_2} \omega_2
}
is a calibration for any $-1 \leq s, t  \leq 1$.  Here,  by $\omega_1$ and $\omega_2$ we mean, as usual, the pull-backs of the K\"ahler forms on $V_1$ and $V_2$, respectively.  Clearly, since we fix $r$, $\Omega$ is a closed two-from.  To understand why $\Omega$ is in fact a calibration, let us pick a point $p$ in $\tilde Y^7$ and define the orthonormal basis $f_i$, $i = 1, 2, \ldots, 7$, for the tangent space $T_p \tilde Y^7$ and the dual basis $e_j$, $j = 1, 2, \ldots, 7$ for $T_p^* \tilde Y^7$.  Since $\omega_i$ are the pull-backs of the K\"ahler forms on $V_i$, we can require
\es{ONBasis}{
e^{\chi + \eta_1} \omega_1 &= e_1 \wedge e_2 + e_3 \wedge e_4 \,, \\
e^{\chi + \eta_2} \omega_2 &=e_5 \wedge e_6 \,, \\
e^{{1 \over 2} \chi - {1 \over 2} \eta_1 - \eta_2} (d \psi + \sigma_1 + \sigma_2) &= e_7 \,,
}
and thus the metric on $\tilde Y^7$ is $ds_{\tilde Y}^2 = \sum_{i = 1}^7 (e_i)^2$.  Now for any two arbitrary orthonormal tangent vectors $u = \sum_{i = 1}^7 u_i f_i$ and $v = \sum_{i = 1}^7 v_i f_i$ in $T_p \tilde Y^7$ we have
\es{CalibONVectors}{
\Omega(u, v)  &=  s (u_1 v_2 - u_2 v_1 + u_3 v_4 - u_4 v_3) + t (u_5 v_6 - u_6 v_5) \\
&\leq \left[ s^2 (u_1^2 + u_2^2 + u_3^2 + u_4^2) + t^2 (u_5^2 + u_6^2) \right]^{1\over 2}
\left[v_1^2 + v_2^2 + v_3^2 + v_4^2 + v_5^2 + v_6^2 \right]^{1\over 2}\\
&\leq \norm{u} \norm{v} = 1 \,,
}
where in the second line we used the Cauchy-Schwarz inequality and in the last line we made use of the fact that $-1 \leq s, t \leq 1$.  Equation \eqref{CalibONVectors} holds for any orthonormal vectors $u, v$ at any point $p$, so $\Omega$ is indeed a calibration.  For a surface ${\cal C}$ in $\tilde Y^7$ we therefore have
\es{VolumeInequality}{
\Vol({\cal C}) \geq e^{\chi + \eta_1} \left| \int_{\cal C} \omega_1 \right| + e^{\chi + \eta_2} \left| \int_{\cal C} \omega_2\right| \,.
}
In obtaining \eqref{VolumeInequality} we chose $s$ and $t$ to be $\pm 1$ in such a way that the bound we got would be as restrictive as possible.

Combining \eqref{VolumeInequality} with \eqref{LiftableConditionSimp}, we obtain\footnote{In the case of $AdS_5 \times T^{1, 1}$ a similar inequality was proven in \cite{Arean:2004mm} using an explicit parameterization of two-cycles.}
\es{VolumeIneqSimpAppendix}{
\Vol({\cal C}) \geq e^{\chi} \left( e^{\eta_1} + e^{\eta_2} \right) \left| \beta \int_D \omega_2 \right|
=  e^{\chi} \left( e^{\eta_1} + e^{\eta_2} \right) \left| \int_{\cal C} \omega_2 \right|\,.
}
This inequality is saturated when both inequalities in \eqref{CalibONVectors} are saturated at every point $p$ of ${\cal C}$, $u$ and $v$ being an orthonormal basis for the tangent space to ${\cal C}$ at $p$.  The first inequality in \eqref{CalibONVectors} is saturated when the projection of ${\cal C}$ to $V_1$ is given by a holomorphic ($s = 1$) or anti-holomorphic ($s = -1$) surface and the projection to $V_2$ is also given by a holomorphic ($t = 1$) or anti-holomorphic ($t = -1$) surface.  The second inequality in \eqref{CalibONVectors} is satisfied when the tangent space to ${\cal C}$ is ``horizontal,'' meaning intuitively that ${\cal C}$ does not ``move'' in the fiber direction.  Only very special surfaces satisfy these two conditions.  That said, two such surfaces are the one given in \eqref{CM111} in the case of $M^{1, 1, 1}$ and the one given in \eqref{C2Q111} in the case of $Q^{1, 1, 1}$;  a direct computation of the volumes of these surfaces shows that they indeed saturate \eqref{VolumeIneqSimp}.

\section{Reduction to type IIA and T-duality}
\label{APPENDIXREDUCE}

In this section we reduce the M-theory background \eqref{11dMetric}--\eqref{F4M2} to type IIA along the $x^2$ direction and then T-dualize to type IIB along the $x^1$ direction.  The type IIA string frame metric is
\es{IIAMetric}{
ds_{\rm IIA}^2 &= e^{-{21 \over 4} \chi} {r \over L} \left[-g e^{-w} dt^2 + {r^2 \over L^2} (dx^1)^2 + {dr^2 \over g} \right] \\
{}&+ 4 L^2 e^{-{3 \over 4} \chi} {r \over L} \left[ e^{\eta_1} ds_{V_1}^2 + e^{\eta_2} ds_{V_2}^2
+ e^{-4 \eta_1 - 2 \eta_2} (d\psi + \sigma_1 + \sigma_2)^2 \right] \,.
}
The dilaton is given by
\es{IIADilaton}{
\Phi_{\rm IIA} = -{21 \chi \over 8} + {3 \over 2} \log {r\over L} \,.
}
The NS-NS three-form flux is
\es{IIAH3}{
H_3^{\rm IIA} = 3 e^{-{1 \over 2} w - {21 \over 2} \chi} {r^2 \over L^3} dt \wedge dx^1 \wedge dr \,.
}
Out of the R-R forms, only $F_4$ is non-vanishing:
\es{F4IIA}{
F_4^{\rm IIA} = - 8 Q e^{-{1 \over 2} w - {3 \over 2} \chi} {L^3 \over r^2}
dt \wedge dr \wedge \left(e^{2 \eta_1} \omega_1 - 2 e^{2 \eta_2} \omega_2 \right) \,.
}

The type IIB background we obtain has only $F_5$ flux.  In string frame, the metric is
\es{IIBMetric}{
ds_{\rm IIB}^2 &=  e^{- {21\over 4} \chi} {r \over L} \left[-g e^{-w} dt^2 + {dr^2 \over g} \right]
+ e^{{21 \over 4} \chi} {L^3 \over r^3}  \left(dx^1 - P(r) dt \right)^2 \\
{}&+ 4 L^2 e^{-{3 \over 4} \chi} {r \over L} \left[e^{\eta_1} ds_{V_1}^2 + e^{\eta_2} ds_{V_2}^2
+ e^{-4 \eta_1 - 2 \eta_2} (d\psi + \sigma_1 + \sigma_2)^2 \right] \,,
}
where the function $P(r)$ satisfies
\es{Pprime}{
P'(r) = 3 e^{-{1 \over 2} w - {21 \over 2} \chi} {r^2 \over L^3} \,.
}
The self-dual five-form can be written as
\es{F5}{
F_5^{\rm IIB} = 8 Q e^{-{1 \over 2} w - {3 \over 2} \chi} {L^3 \over r^2}
dt \wedge dx^1 \wedge dr \wedge \left( e^{2 \eta_1} \omega_1 - 2 e^{2 \eta_2} \omega_2 \right)
 -  64 Q L^4 *_Y \omega\,,
}
where $*_Y \omega$ was defined in \eqref{HodgeDuals}.

\bibliographystyle{ssg}
\bibliography{wrappedM2}

\end{document}